\documentclass[aps,prl,reprint,twocolumn]{revtex4-2}

\usepackage{amsmath} 
\usepackage{graphicx}
\usepackage{dcolumn}
\usepackage{bm}%
\usepackage{braket}
\usepackage{physics}
\usepackage{amsmath}  
\usepackage{amsthm}   
\usepackage{amssymb}  
\usepackage{xcolor} 

\theoremstyle{plain}
\newtheorem{lemma}{Lemma}

\theoremstyle{remark}
\newtheorem{remark}{Remark}
\begin{document}

\title{Quantum Speed Limit under Calibration Uncertainty}

 \author{Salman Sajad Wani}
 \affiliation{Qatar Center for Quantum Computing, College of Science and Engineering, Hamad Bin Khalifa University, Doha, Qatar} 
\author{Saif Al-Kuwari}
\affiliation{Qatar Center for Quantum Computing, College of Science and Engineering, Hamad Bin Khalifa University, Doha, Qatar}

\begin{abstract}
Standard quantum speed limits presuppose exactly known parameters, overestimating operational speed under calibration uncertainty. We introduce a projected speed limit based on the quantum Fisher information that profiles out these nuisance parameters on a quotient manifold. We derive constructive bounds for general Markovian evolution using sensitivity equations. Applying this to Jaynes--Cummings sensors, we obtain explicit detuning tolerances and quantify speed limits arising from field-dependent Purcell loss. This framework turns geometric bounds into concrete design rules for calibration and interrogation time.
\end{abstract}

\maketitle
\section{Introduction}
Quantum speed limit (QSL) quantify the minimal time for a quantum state to evolve into a distinguishable state. The earliest bounds, due to Mandelstam--Tamm and Margolus--Levitin, relate the orthogonalization time to the energy variance and, respectively, to the mean energy above the ground state \cite{mandelstam1945,anandan1990,margolus1998}. Geometric treatments recast these results in the Bures/Uhlmann geometry and use the symmetric logarithmic derivative (SLD) quantum Fisher information (QFI) metric to fix the line element \cite{bures1969,uhlmann1976,braunstein1994,petz1996}. Subsequent work extended this framework to general open-system dynamics, yielding QSLs expressed directly in terms of the instantaneous statistical speed for completely positive trace-preserving evolutions \cite{taddei2013,delcampo2013,deffner2013}. {Related developments also exploited the nonuniqueness of contractive distinguishability metrics to construct generalized geometric QSL families for unitary and nonunitary dynamics, including bounds that can be tighter than the conventional QFI/Bures choice \cite{pires2016}. For open dynamics,  studies have emphasized complementary desiderata such as tightness, robustness under composition and mixing, and computational or experimental feasibility \cite{campaioli2019}. Related work also sharpened the operational interpretation of quantum speed limits, including in open-system settings \cite{OConnorGuarnieriCampbell2021,ShaoLiuZhangYuanLiu2020}. More recent Liouville-space formulations have derived exact and inexact QSL families for completely positive trace-preserving dynamics and clarified attainability for time-optimal CPTP evolutions \cite{srivastav2025}.}
These developments have deepened the link between QSLs, quantum estimation theory, and monotone information metrics \cite{helstrom1976,petz1996,petzSudar1996}. They also underpin many analyses of quantum metrology, control, and dissipation \cite{giovannetti2011,lindblad1976,gks1976}.

{A common feature across the standard Bures/QFI setting, generalized geometric families, and more recent open-system and CPTP formulations is the assumption of a fixed, known generator \cite{taddei2013,delcampo2013,deffner2013,pires2016,campaioli2019,srivastav2025}.} Operationally, distinguishability along laboratory time is therefore confounded by nuisance directions in parameter space: an experimenter can partially compensate for changes in time by co-varying these parameters within their estimation uncertainty. Consequently, the physical Bures speed computed at a fixed generator generally overestimates the operational speed relevant for distinguishing states modulo nuisance reparametrizations.

In this paper we formulate a nuisance-aware projected quantum speed limit. We treat time and nuisance parameters jointly within the multiparameter SLD-QFI and project the Bures metric onto the time direction via a Schur complement \cite{coxReid1987,kay1993,Zhang2005}. This construction yields an effective information $F_{\mathrm{eff}}$ and an associated projected speed $v_{\mathrm{quo}}=\tfrac12\sqrt{F_{\mathrm{eff}}}$. This speed never exceeds the standard Bures speed and remains invariant under reparametrizations of the nuisance manifold. Crucially, $F_{\mathrm{eff}}$ coincides with the partial SLD Fisher information, which sets the ultimate precision limit when nuisance parameters are profiled out \cite{Suzuki2020Nuisance,SidhuKok2020Multiparameter}. In contrast to standard formulations \cite{taddei2013,delcampo2013,deffner2013}, which bound evolution for a frozen generator, our framework quantifies operational distinguishability modulo nuisances. We implement this for general Markovian dynamics using GKSL sensitivity equations \cite{lindblad1976,gks1976,Pazy1983Semigroups,EngelNagel2000Semigroups} and analyze two paradigmatic sensors: a unitary Jaynes-Cummings (JC) model and a dispersive JC cavity with loss.  For the examples studied here, the nuisance penalty is more pronounced in the dispersive open JC sensor, since the field controls both the coherent cavity-frequency shift and the effective decay rate. In both regimes, the projected QSL converts abstract geometric limits into concrete design rules. One example is the tolerance $|\Delta|t\lesssim 0.3$ for retaining $99\%$ of the physical speed.
\section{Formalism}
{We work in the Bures--SLD geometry of quantum states. The
Bures metric, induced by the Uhlmann fidelity, provides the local statistical
distance on the space of density operators
\cite{bures1969,uhlmann1976,braunstein1994,petz1996}. The corresponding
symmetric logarithmic derivative (SLD) operators $L_\mu$ are defined by
\begin{equation}
\partial_\mu \rho=\tfrac12(\rho L_\mu+L_\mu\rho),
\end{equation}
and the associated SLD quantum Fisher information matrix (QFIM)
$F_{\mu\nu}$ determines the local line element
\begin{equation}
dL^{2}=\tfrac14 F_{\mu\nu}\,dx^\mu dx^\nu .
\end{equation}
For any parameter direction, this quadratic form gives the local sensitivity of
the state to infinitesimal displacements; for time evolution, its square root is
the instantaneous statistical speed that underlies geometric quantum speed
limits \cite{braunstein1994,taddei2013,deffnerCampbell2017}.} We take the coordinate vector $x^{\mu}=(t,\bm\lambda)$ to include both laboratory time and the nuisance parameters $\bm\lambda$ (e.g., detunings, loss rates). In our setting, $\bm\lambda$ are not targets of joint estimation but sources of uncertainty. {Throughout this work, the nuisance parameters $\bm\lambda$ are treated as
run-wise fixed but imperfectly known calibration parameters that label the
generator of the dynamics. They are therefore distinct from explicitly
time-dependent stochastic fluctuations, which belong to a different modeling
layer and are not profiled by the present Schur-complement construction.} In this multiparameter setting, the QFIM takes the block form
\begin{equation}
F =
\begin{pmatrix}
F_{tt} & \bm f^{\top} \\
\bm f  & F_{\lambda\lambda}
\end{pmatrix},
\qquad
\bm f := (F_{t\alpha})_\alpha,\;
F_{\lambda\lambda} := (F_{\alpha\beta})_{\alpha,\beta}.
\end{equation}
Standard ``physical'' QSLs are obtained by fixing $\bm\lambda$ and using $F_{tt}$ alone. Operationally, however, an experimenter can re-tune $\bm\lambda$ to reduce distinguishability along nuisance directions. Thus the relevant speed is set not by $F_{tt}$, but by the information component that is orthogonal to the nuisance manifold.

To isolate this component, we minimize $dL^{2}$ over $d\bm\lambda$ at fixed $dt$, profiling the Bures quadratic form in analogy with profile likelihood \cite{coxReid1987,kay1993,AmariNagaoka2000InformationGeometry}. Solving the normal equations yields the Schur complement
\begin{equation}
F_{\mathrm{eff}}
:= F_{tt} - \bm f^{\top} F_{\lambda\lambda}^{-1} \bm f ,
\end{equation}
where $F_{\lambda\lambda}^{-1}$ denotes the Moore--Penrose pseudoinverse on the estimable nuisance subspace \cite{benisrael2003,Zhang2005}. {
Geometrically, this Schur complement is the squared norm of the time tangent
after orthogonal projection away from the nuisance directions in the Bures--SLD geometry. It therefore removes the distinguishability attributable to nuisance
reparametrizations. Equivalently, the profiled quadratic form induces the local
quotient metric obtained by identifying tangent variations that differ only by
nuisance components:
$ dL_{\mathrm{quo}}^{2}=1/4 F_{\mathrm{eff}}\,dt^{2}.$}  $F_{\mathrm{eff}}$ satisfies $0 \le F_{\mathrm{eff}} \le F_{tt}$ and is invariant under smooth reparametrizations of $\bm\lambda$. When $F$ is invertible, $F_{\mathrm{eff}} = ((F^{-1})_{tt})^{-1}$, recovering the standard profiled information. Rank-deficient states (including pure states) are handled by restricting all constructions to $\operatorname{supp}(\rho)$ \cite{helstrom1976,safranek2017}.

In the SLD formalism, $F_{\mathrm{eff}}$ coincides with the {partial} quantum Fisher information for time. Projecting $L_t$ orthogonally to the span of nuisance scores $\{L_\alpha\}$ via the inner product $\langle\!\langle A,B\rangle\!\rangle_\rho := \tfrac12\Tr[\rho\{A,B\}]$ yields the efficient score $\widetilde L_t$, whose norm is
\begin{equation}
\langle\!\langle \widetilde L_t,\widetilde L_t\rangle\!\rangle_\rho
= F_{tt} - \bm f^\top F_{\lambda\lambda}^{-1}\bm f
= F_{\mathrm{eff}}.
\end{equation}
Thus, the projected metric coefficient sets the ultimate precision limit $\mathrm{Var}(\hat t)\ge 1/(N F_{\mathrm{eff}})$ attainable when nuisance parameters are profiled out \cite{Suzuki2020Nuisance,SidhuKok2020Multiparameter}.

To embed this geometry into Markovian dynamics, we consider trajectories generated by the GKSL master equation
\begin{equation}\label{eq:GKSL}
\partial_t \rho(t,\bm\lambda)
= \mathcal{L}_{\bm\lambda}[\rho]
:= -\tfrac{i}{\hbar}[H(\bm\lambda),\rho]
+ \sum_k \mathcal{D}[F_k(\bm\lambda)]\rho .
\end{equation}
The time derivative entering the QFIM is $\partial_t\rho = \mathcal{L}_{\bm\lambda}[\rho]$, while the nuisance derivatives $\rho'_\alpha := \partial_{\lambda^\alpha}\rho$ satisfy the linear sensitivity equation
\begin{equation}
\partial_t \rho'_\alpha(t)
= (\partial_{\lambda^\alpha}\mathcal{L}_{\bm\lambda})[\rho(t)]
+ \mathcal{L}_{\bm\lambda}[\rho'_\alpha(t)].
\end{equation}
This equation allows $F_{\mathrm{eff}}(t)$ to be evaluated efficiently along any trajectory using the Duhamel formula \cite{Pazy1983Semigroups,EngelNagel2000Semigroups}. {
At each fixed time, the profiling step remains explicit: one solves the normal
equations
$F_{\lambda\lambda}\,d\bm\lambda_{*}=-\,\bm f\,dt$
and substitutes the result into the quadratic form, recovering the same Schur
complement as above. The effective information defines the instantaneous projected speed
$v_{\mathrm{quo}}(t):=\frac12\sqrt{F_{\mathrm{eff}}(t)}$.
Its time integral is the quotient length of the realized path, while the global
minimization is already contained in the quotient geodesic distance
$L_{\mathrm{quo}}([\rho_0],[\rho_\tau])$.}
Integrating this speed yields the projected quantum speed limit
\begin{equation}
\tau \;\ge\;
\frac{L_{\mathrm{quo}}\big([\rho_0],[\rho_\tau]\big)}{\bar v_{\mathrm{quo}}},
\qquad
\bar v_{\mathrm{quo}}
:= \frac{1}{\tau}\int_0^\tau \tfrac12\sqrt{F_{\mathrm{eff}}(t)}\,dt ,
\end{equation}
where $L_{\mathrm{quo}}$ is the geodesic distance between equivalence classes $[\rho]$ on the quotient manifold. In the absence of nuisances ($F_{t\alpha}=0$), this reduces to standard QFI-based bounds \cite{taddei2013,delcampo2013,deffner2013}. For unitary dynamics, it yields a Mandelstam-Tamm--type bound governed by an effective variance $\Delta_{\mathrm{eff}}$ \cite{mandelstam1945,anandan1990} (see Appendix~\ref{app:formalism}).
{The projected QSL is saturated when the realized evolution,
viewed in the quotient geometry obtained after profiling out nuisance
directions, is a minimizing geodesic between the endpoint equivalence classes.
Equivalently,
\begin{equation}
L_{\rm quo}\big([\rho_0],[\rho_\tau]\big)
=
\int_0^\tau \frac12\sqrt{F_{\rm eff}(t)}\,dt.
\end{equation}
When $F_{t\alpha}=0$, one has $F_{\rm eff}=F_{tt}$, so the projected-QSL
denominator coincides with the standard geometric denominator. If the realized
endpoint also attains the quotient optimization over the admissible nuisance
values, then the quotient endpoint distance coincides with the physical Bures
distance for that run, and the saturation question reduces to the usual one for
the underlying dynamics.}

\section{Unitary example: Jaynes--Cummings sensor.}
We first illustrate the projected QSL in a setting where estimation and dynamics are statistically correlated: a Jaynes--Cummings (JC) qubit--cavity sensor for a magnetic field $B$ \cite{jaynes1963,gardiner2004,carmichael1993}. Within the rotating--wave approximation and restricted to the single--excitation manifold $\{\ket{e,0},\ket{g,1}\}$, the dynamics is generated by the traceless Hamiltonian
\begin{equation}
  H'(B)=g\,\tau_x+\frac{\Delta(B)}{2}\,\tau_z,\qquad
  \Delta(B)=\omega_q+\gamma B-\omega_c,
\end{equation}
with associated Rabi frequency $\Omega=\sqrt{\Delta^2+4g^2}$. We initialize in $\ket{\psi_0}=\ket{e,0}$ and treat the field $B$, through the detuning $\Delta(B)$, as a nuisance parameter.

For a pure state $\ket{\psi(x)}=U(x)\ket{\psi_0}$, the QFIM is given by the covariance of the local generators, $F_{\mu\nu}=4\,\mathrm{Cov}_{\psi_0}(G_\mu,G_\nu)$ \cite{braunstein1994}. Here the time generator is $G_t=H'(B)$, while the field generator $G_B(t)$ is dynamical because $[H',\partial_B H']\neq 0$. An explicit evaluation of the Heisenberg equations (see Appendix~\ref{app:JC-unitary}) yields
\begin{subequations}
\begin{align}
  F_{tt} &= 4g^2, \\
  F_{tB}(t) &= \frac{4g^2\gamma\,\Delta}{\Omega^2}
              \left(t-\frac{\sin\Omega t}{\Omega}\right),
\end{align}
\end{subequations}
so that the diagonal element $F_{tt}$ reproduces the standard Mandelstam--Tamm speed $v_{\rm phys}=2g$, while the nonzero off--diagonal term $F_{tB}$ captures the correlation between time evolution and field uncertainty.

Eliminating the field uncertainty via the Schur complement $F_{\mathrm{eff}} = F_{tt} - F_{tB}^2/F_{BB}$ yields the effective information
\begin{equation}
  F_{\mathrm{eff}}(t)
  = 4g^2\,
   \frac{4\sin^{4}\!\bigl(\tfrac{\Omega t}{2}\bigr)}
        {\displaystyle
          \Delta^2\!\left(t-\tfrac{\sin\Omega t}{\Omega}\right)^{\!2}
          + 4\sin^{4}\!\bigl(\tfrac{\Omega t}{2}\bigr)}.
\end{equation}
Two regimes are particularly instructive. (i) At exact resonance ($\Delta=0$), the correlation $F_{tB}$ vanishes and the projected speed coincides with the physical speed. (ii) For small detunings and short times ($|\Omega t|\ll 1$), a Taylor expansion gives a quadratic penalty:
\begin{equation}
  \frac{F_{\mathrm{eff}}(t)}{F_{tt}}
   \approx \frac{1}{1+(\Delta t/3)^2}
   \simeq 1-\frac{(\Delta t)^2}{9}.
\end{equation}
Thus, even a modest residual detuning reduces the operational speed $v_{\rm quo}$ below the naive Mandelstam--Tamm value.

\begin{figure}[t]
    \centering
    \includegraphics[width=1.0\linewidth]{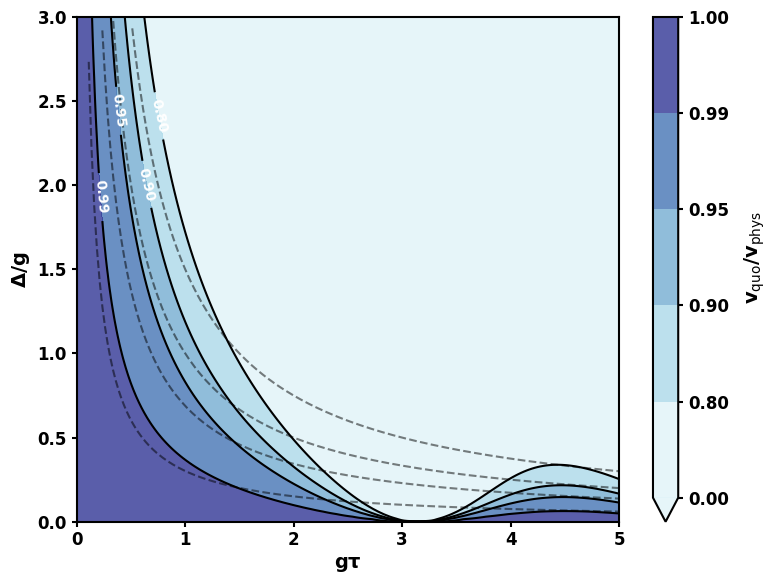}
    \caption{\textbf{Calibration phase diagram for the unitary Jaynes--Cummings sensor.}
    The heatmap shows the ratio of projected to physical speed, $\mathcal{R} = v_{\mathrm{quo}}/v_{\mathrm{phys}}$, as a function of the dimensionless interrogation time $g\tau$ and the normalized detuning $\Delta/g$.
    White solid contours mark the retention levels $R=0.99, 0.95, 0.90,$ and $0.80$.
    Black dashed curves show the analytical tolerance $|\Delta|\tau \approx 3\sqrt{1/R-1}$ derived in Eq.~\eqref{eq:tolerance}.
    The dark blue region marks the admissible $(\Delta,\tau)$ working window in which nuisance profiling has only a small effect on the operational quantum speed.}
    \label{fig:unitary_contours}
\end{figure}

This reduction leads directly to quantitative design rules. If we require the projected speed to retain a fraction $R$ of the physical speed, we obtain the tolerance
\begin{equation}
  \label{eq:tolerance}
  |\Delta|\,t < 3\sqrt{\frac{1}{R}-1}.
\end{equation}
For example, keeping $R=0.99$ requires $|\Delta|t < 0.30$. This inequality shows how tightly the JC sensor must be kept near resonance for the operational QSL to remain close to the QSL computed at fixed generator. Profiling nuisance parameters therefore turns the geometric construction into concrete constraints on calibration effort and on the admissible $(\Delta,\tau)$ working window (see Fig.~\ref{fig:unitary_contours}).

\section{Open example: Dispersive JC sensor with loss.}
We next consider a realistic sensor with noisy, parameter-dependent dynamics: a Jaynes--Cummings cavity coupled to a two-level atom, operated in the dispersive regime ($\Delta \gg g$) \cite{jaynes1963,blais2004}. {Here the dissipation is modeled by a Markovian master equation for fixed $B$,
while the uncertainty in $B$ itself is treated as static calibration uncertainty
during a given run.}
The atom acts as a probe for the field $B$, shifting the cavity frequency via the AC Stark effect and at the same time introducing dissipation.

\begin{figure}[t]
    \centering
    \includegraphics[width=1.0\linewidth]{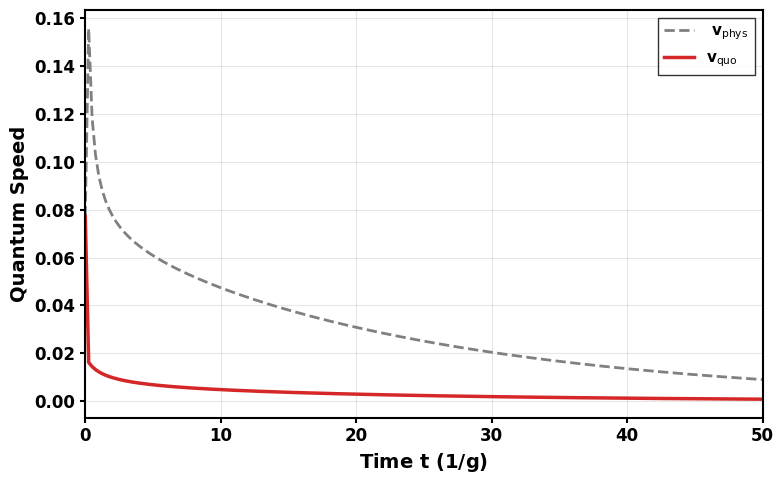}
    \caption{\textbf{Operational speed penalty from Purcell-enhanced loss.} 
    The figure shows the instantaneous physical speed $v_{\mathrm{phys}}$ (gray dashed) and projected speed $v_{\mathrm{quo}}$ (red solid) for a dispersive Jaynes--Cummings sensor with parameters $\Delta=8g$, $\kappa=0.05g$, and $\gamma_1=2g$. 
    Whereas the physical speed remains high because the state continues to decay dissipatively, the projected speed shows a pronounced {nuisance penalty} driven by the statistical correlation between time and the field-dependent decay rates.}
    \label{fig:open_speed}
\end{figure}

By applying a Schrieffer--Wolff transformation \cite{blais2004,schrieffer1966} and tracing out the fast-relaxing atom (see Appendix~\ref{app:JC-open}), we obtain an effective master equation for the cavity mode $\rho_{\rm cav}$:
\begin{equation}
\label{eq:cavity-master}
  \dot{\rho}_{\rm cav}
  = -i\big[\omega_c'(B)\,\tilde{a}^\dagger\tilde{a},\rho_{\rm cav}\big]
    + \kappa_{\rm eff}(B)\,\mathcal{D}[\tilde{a}]\,\rho_{\rm cav}.
\end{equation}
{
In the dispersive regime, the Schrieffer--Wolff transformation removes the
off-resonant atom--cavity exchange term order by order in $g/\Delta(B)$ and
leaves a dressed cavity mode whose resonance is shifted by virtual atomic
excitations. The same dressing admixes a small atomic component into the
effective cavity excitation, so atomic relaxation is inherited by the cavity as
an additional decay channel, yielding the field-dependent Purcell contribution
to $\kappa_{\rm eff}(B)$ \cite{purcell1946}. Hence the same parameter $B$
controls both the coherent signal, carried by the field-pulled frequency
$\omega_c'(B)$, and the dissipative envelope, through the effective decay rate
$\kappa_{\rm eff}(B)=\kappa+\delta\kappa(B)$. This joint dependence is the
physical origin of the trade-off in Fig.~\ref{fig:open_speed}: increasing the
dispersive response simultaneously strengthens the loss channel, which makes the
passage of time more easily mimicked by a change in the field-dependent decay
rate and therefore increases the nuisance penalty in the projected speed.
}
For an initial superposition state, the evolved density matrix takes the Bloch form $\rho_{\rm cav}(t) = \tfrac{1}{2}(\mathbb{I} + \vec{S}\cdot\bm{\sigma})$. 
The Bloch components then evolve as
\begin{equation}
\begin{aligned}
  S_x &= \sin(2\theta)\cos[\omega_c'(B)t]\,e^{-\kappa_{\rm eff}(B)t/2},\\
  S_y &= -\sin(2\theta)\sin[\omega_c'(B)t]\,e^{-\kappa_{\rm eff}(B)t/2},\\
  S_z &= 2\sin^2\theta\,e^{-\kappa_{\rm eff}(B)t}-1.
\end{aligned}
\end{equation}

Accordingly, the parameter $B$ enters both the oscillation frequency and the exponential decay envelope. 
The effective information 
$F_{\rm eff}(t) = F_{tt} - F_{tB}^2/F_{BB}$ 
quantifies the balance between coherent phase accumulation and irreversible amplitude decay.

These effects are summarized in Fig.~\ref{fig:open_speed}. 
The physical speed $v_{\rm phys}$ simply tracks the continuous decay of the state, whereas the projected speed $v_{\rm quo}$ falls off rapidly and reveals a growing {nuisance penalty}. 
This divergence arises because photon loss is statistically confounded with uncertainty in $B$: an experimenter cannot tell whether a weaker signal is caused by the passage of time or by a change in the Purcell rate.

To bound the evolution, we compute the quotient fidelity by maximizing over the calibration window $\mathcal{B}$:
\begin{equation}
  \Theta_{\rm quo}(\tau)
  := \arccos\!\Big(\sup_{B\in\mathcal{B}}\sqrt{F(\rho_0,\rho_{\rm cav}(\tau;B))}\Big).
\end{equation}
Combining this with the averaged projected speed gives the open-system bound $\tau \ge \Theta_{\rm quo}/\bar v_{\rm quo}$. 
This result identifies a finite optimal interrogation time, before the nuisance penalty dominates. 
In this way, the geometric limit becomes a concrete design rule for dispersive readout \cite{blais2004,gambetta2006}.

\section{Conclusion}
We introduced a nuisance-aware quantum speed limit on the quotient manifold of physical states, which identifies trajectories that differ only by unobservable reparametrizations of the generator. 
Projecting the multiparameter Bures metric via the Schur complement \cite{coxReid1987,kay1993} yields an operational speed $v_{\mathrm{quo}}$, which provides a lower bound on distinguishability in the presence of calibration uncertainty. 
Unlike standard QSLs that assume a fixed generator \cite{taddei2013,delcampo2013}, this framework captures the statistical confounding between time evolution and nuisance parameters.

By embedding Markovian dynamics into this geometry, we turn abstract information-geometric limits into concrete design rules. 
As demonstrated for the unitary and dissipative Jaynes--Cummings sensors, the projected QSL sets explicit experimental tolerances—such as the detuning limit $|\Delta|t \lesssim 0.3$ and the Purcell-penalty cutoff—needed to approach fundamental physical bounds. 
This shift from kinematic constraints to operational benchmarks enables nuisance-profiled metrology in complex architectures and ties idealized geometric theorems to the noisy, uncertain behavior of real quantum hardware.

 \bibliography{name1}

\clearpage
\begin{widetext}
\setcounter{section}{0}
\setcounter{equation}{0}
\setcounter{figure}{0}
\setcounter{table}{0}

\makeatletter
\renewcommand{\thesection}{\Alph{section}}
\renewcommand{\theequation}{\thesection\arabic{equation}}
\renewcommand{\thefigure}{\thesection\arabic{figure}}
\renewcommand{\thetable}{\thesection\arabic{table}}
\makeatother

\refstepcounter{section}
\setcounter{equation}{0}
\setcounter{figure}{0}
\setcounter{table}{0}
\section*{Appendix \thesection: Detailed formalism --- Bures geometry, Schur complement, and GKSL dynamics}
\label{app:formalism}

In this section we derive, in full detail, the geometric quantities and constructions used in the main text:
the Bures metric and SLD quantum Fisher information matrix (QFIM), the Schur-complement projection that
profiles out nuisance parameters, and the embedding of Markovian dynamics via GKSL sensitivity equations.
This provides the rigorous backbone for the projected (quotient) quantum speed limit used in Sec.~II of
the main text.

Throughout we work on a finite-dimensional Hilbert space $\mathcal{H}$.
Parameters are real and collected as
\begin{equation}
  x^\mu = (t,\lambda^1,\dots,\lambda^{m-1}) \in \mathbb{R}^m,
\end{equation}
with $C^1$ dependence of the state family $\rho(x)$ and of the GKSL data
$H(\bm\lambda)$, $\{F_k(\bm\lambda)\}$.

\subsection{Setting and GKSL dynamics}
\label{subsec:SM-GKSL-setting}

For each fixed $\bm\lambda$, the laboratory time evolution is generated by a
time-independent GKSL (Lindblad) generator
\begin{equation}
  \partial_t \rho(t,\bm\lambda)
  = \mathcal{L}_{\bm\lambda}[\rho]
  := -\frac{i}{\hbar}[H(\bm\lambda),\rho]
  + \sum_k \left(F_k \rho F_k^\dagger - \frac12 \{F_k^\dagger F_k,\rho\}\right).
  \label{eq:SM-GKSL}
\end{equation}
See Refs.~\cite{GoriniKossakowskiSudarshan1976,lindblad1976,BreuerPetruccione2002}.
Normalization $\Tr\rho(t,\bm\lambda)=1$ implies that for any tangent $\delta\rho$ one
has $\Tr(\delta\rho)=0$.

Our goal in this section is to start from the Uhlmann fidelity, derive the Bures
metric and the SLD--QFIM, project away nuisance directions by a Schur complement
to obtain an effective (profiled) information $F_{\rm eff}$, and then combine this
with the GKSL dynamics \eqref{eq:SM-GKSL} to obtain a projected (quotient) quantum
speed limit.

\subsection{From Uhlmann fidelity to the Bures metric}
\label{subsec:SM-Uhlmann}

The Uhlmann fidelity between two density operators $\rho$ and $\sigma$ is
defined as
\begin{equation}
  F_U(\rho,\sigma) := \left(\Tr\sqrt{\sqrt{\rho}\,\sigma\,\sqrt{\rho}}\right)^2.
\end{equation}
For nearby states $\sigma=\rho+\delta\rho$, the fidelity admits the expansion
\begin{equation}
  F_U(\rho,\rho+\delta\rho)
  = 1 - g_\rho(\delta\rho,\delta\rho) + o(\|\delta\rho\|^2),
  \label{eq:SM-Uhlmann-expansion}
\end{equation}
which {defines} the Bures metric tensor $g_\rho$
\cite{uhlmann1976,bures1969,hubner1992,Dittmann1999}.

{To make this explicit, let $\rho$ be full rank (the rank-deficient case is handled below in the subsection ``Rank-deficient states and singular nuisance
blocks''), with spectral decomposition}equation A
\begin{equation}
  \rho = \sum_i p_i \ket{i}\bra{i}, \qquad p_i>0.
\end{equation}
Set
\begin{equation}
  \sigma = \rho + \delta\rho, \qquad
  \mathcal{A} := \sqrt{\rho}\,\sigma\,\sqrt{\rho}
  = \rho^2 + \sqrt{\rho}\,\delta\rho\,\sqrt{\rho},
\end{equation}
and define the positive square root
\begin{equation}
  \mathcal{M} := \sqrt{\mathcal{A}}.
\end{equation}
We expand $\mathcal{M}$ in powers of $\delta\rho$:
\begin{equation}
  \mathcal{M} = \rho + X + Y + O(\|\delta\rho\|^3),
\end{equation}
where $X=O(\delta\rho)$ collects linear terms and $Y=O(\delta\rho^2)$ collects
quadratic terms. The defining identity $\mathcal{M}^2=\mathcal{A}$ then yields
two Sylvester equations:
\begin{align}
  \rho X + X \rho &= \sqrt{\rho}\,\delta\rho\,\sqrt{\rho},
  \label{eq:SM-L1}\\
  \rho Y + Y \rho &= -X^2.
  \label{eq:SM-Q1}
\end{align}
See, e.g.,~\cite{hubner1992,Dittmann1999,HornJohnson2013}.

In the eigenbasis of $\rho$, the solution of \eqref{eq:SM-L1} is
\begin{equation}
  X_{ij} = \frac{\sqrt{p_i p_j}}{p_i + p_j} (\delta\rho)_{ij},
  \qquad
  \Tr X = \frac12 \Tr(\delta\rho) = 0,
  \label{eq:SM-Xij}
\end{equation}
where we used $\Tr(\delta\rho)=0$. To obtain $\Tr Y$, we work with
\eqref{eq:SM-Q1}. Taking the trace in the eigenbasis,
\begin{align}
  \Tr Y
  &= -\frac12 \Tr(\rho^{-1}X^2)
   = -\frac12 \sum_{i,j}\frac{p_j}{(p_i+p_j)^2}\,|(\delta\rho)_{ij}|^2.
\end{align}
Using symmetry under $i\leftrightarrow j$,
\begin{align}
  \sum_{i,j}\frac{p_j}{(p_i+p_j)^2}\,|(\delta\rho)_{ij}|^2
  &= \frac12 \sum_{i,j}
     \frac{p_i+p_j}{(p_i+p_j)^2}\,|(\delta\rho)_{ij}|^2 \nonumber\\
  &= \frac12 \sum_{i,j}\frac{|(\delta\rho)_{ij}|^2}{p_i+p_j},
\end{align}
so that
\begin{equation}
  \Tr Y = -\frac14\sum_{i,j} \frac{|(\delta\rho)_{ij}|^2}{p_i+p_j}.
  \label{eq:SM-TrY}
\end{equation}

Next, expand the square root of the fidelity.
By definition,
\begin{equation}
  \sqrt{F_U(\rho,\sigma)} = \Tr\sqrt{\mathcal{A}} = \Tr\mathcal{M}.
\end{equation}
Using $\Tr\rho=1$ and $\Tr X=0$, we find
\begin{equation}
  \sqrt{F_U(\rho,\rho+\delta\rho)}
  = 1 + \Tr Y + O(\|\delta\rho\|^3)
  = 1 -\frac14\sum_{i,j} \frac{|(\delta\rho)_{ij}|^2}{p_i+p_j}
    +o(\|\delta\rho\|^2),
  \label{eq:SM-sqrtF}
\end{equation}
where we used \eqref{eq:SM-TrY}. Since $F_U = (\sqrt{F_U})^2$, the
second-order expansion of $F_U$ is
\begin{equation}
  F_U(\rho,\rho+\delta\rho)
  = 1 -\frac12\sum_{i,j}\frac{|(\delta\rho)_{ij}|^2}{p_i+p_j}
    + o(\|\delta\rho\|^2).
\end{equation}
Comparing with the defining expansion \eqref{eq:SM-Uhlmann-expansion}, we obtain
the explicit Bures metric tensor
\begin{equation}
  g_\rho(\delta\rho,\delta\rho)
  = \frac12\sum_{i,j}\frac{|(\delta\rho)_{ij}|^2}{p_i+p_j},
  \label{eq:SM-g-explicit}
\end{equation}
as given, e.g., in Refs.~\cite{hubner1992,petzSudar1996,BengtssonZyczkowski2006}.

\subsection{SLD representation and the SLD--QFIM}
\label{subsec:SM-SLD-QFIM}

We now rewrite the metric \eqref{eq:SM-g-explicit} in terms of the symmetric
logarithmic derivative (SLD). For any tangent $\delta\rho$, the SLD
$L(\delta\rho)$ is defined by the Sylvester equation
\begin{equation}
  \delta\rho = \frac12(\rho L + L\rho)
  \quad\Longleftrightarrow\quad
  L = J_\rho^{-1}(\delta\rho),\qquad
  J_\rho(X):=\frac12(\rho X + X\rho).
  \label{eq:SM-SLD-def}
\end{equation}
In the eigenbasis of $\rho$,
\begin{equation}
  L_{ij} = \frac{2(\delta\rho)_{ij}}{p_i+p_j}.
  \label{eq:SM-Lij}
\end{equation}
Substituting \eqref{eq:SM-Lij} into \eqref{eq:SM-g-explicit} and simplifying,
one obtains
\begin{align}
  g_\rho(\delta\rho,\delta\rho)
  &= \frac12\sum_{i,j}\frac{|(\delta\rho)_{ij}|^2}{p_i+p_j} \nonumber\\
  &= \frac12\sum_{i,j}\frac{1}{p_i+p_j}
     \left|\frac{p_i+p_j}{2}L_{ij}\right|^2 \nonumber\\
  &= \frac18\sum_{i,j}(p_i+p_j)|L_{ij}|^2 \nonumber\\
  &= \frac14\sum_{i,j} p_i |L_{ij}|^2
   = \frac14 \Tr(\rho L^2).
  \label{eq:SM-g-SLD}
\end{align}
Here we used symmetry under $i\leftrightarrow j$ and the Hermiticity of $L$.
Equation \eqref{eq:SM-g-SLD} is the standard SLD representation of the Bures
metric \cite{helstrom1976,braunstein1994,petz1996}.

We now introduce coordinates $x^\mu=(t,\lambda^1,\dots,\lambda^{m-1})$ on the
parameter manifold and expand
\begin{equation}
  \delta\rho = \partial_\mu\rho\,dx^\mu.
\end{equation}
Define the coordinate SLD operators $L_\mu$ by
\begin{equation}
  \partial_\mu\rho
  = \frac12(\rho L_\mu + L_\mu\rho),
  \qquad
  L_\mu := J_\rho^{-1}(\partial_\mu\rho).
  \label{eq:SM-coord-SLD}
\end{equation}
By bilinearity of $g_\rho$, the Bures line element can be written as
\begin{align}
  dL^2
  &= g_\rho(\delta\rho,\delta\rho)
   = \frac14 \Tr\big[\rho (\delta L)^2\big] \nonumber\\
  &= \frac14 \Tr\big[\rho(L_\mu L_\nu)\big]\,dx^\mu dx^\nu,
\end{align}
where we used $\delta L = L_\mu dx^\mu$. This identifies the SLD QFIM
$F_{\mu\nu}$ via
\begin{equation}
  dL^2 = \frac14 F_{\mu\nu}\,dx^\mu dx^\nu,
  \qquad
  F_{\mu\nu} := \frac12\Tr\!\big[\rho(L_\mu L_\nu + L_\nu L_\mu)\big].
  \label{eq:SM-Bures-QFIM}
\end{equation}
Since the SLDs are Hermitian, the SLD--QFIM can equivalently be written as the
real part of the unsymmetrized trace:
\begin{equation}
  F_{\mu\nu}
  = \frac12\Tr\!\big[\rho(L_\mu L_\nu + L_\nu L_\mu)\big]
  = \Re\,\Tr(\rho L_\mu L_\nu).
\end{equation}
In particular, when $\Tr(\rho L_\mu L_\nu)$ is real for the parameter pair
under consideration, the real-part symbol may be omitted.

Using \eqref{eq:SM-SLD-def} and \eqref{eq:SM-coord-SLD}, one also obtains a
computational expression that is convenient for open-system dynamics:
\begin{equation}
  F_{\mu\nu}
  = 2\,\Tr\!\big[(\partial_\mu\rho)\,J_\rho^{-1}(\partial_\nu\rho)\big].
  \label{eq:SM-QFIM-J}
\end{equation}
Equations \eqref{eq:SM-Bures-QFIM} and \eqref{eq:SM-QFIM-J} are the standard
SLD--QFIM formulas \cite{helstrom1976,braunstein1994,petz1996,BengtssonZyczkowski2006}.

\subsection{Projecting nuisance directions: Schur complement}
\label{subsec:SM-Schur}

We now treat time $t$ and the nuisance parameters $\bm\lambda$ on equal
footing as coordinates of the parameter manifold, and then project out the
nuisance directions.

Split the indices $\mu=(t,\alpha)$ and block the QFIM as
\begin{equation}
  F = \begin{pmatrix}
    F_{tt} & \bm f^\top\\
    \bm f & F_{\lambda\lambda}
  \end{pmatrix},
  \qquad
  \bm f := (F_{t\alpha})_\alpha,\quad
  F_{\lambda\lambda} := (F_{\alpha\beta})_{\alpha,\beta}.
  \label{eq:SM-QFIM-block}
\end{equation}
The Bures line element \eqref{eq:SM-Bures-QFIM} can then be written as the
quadratic form
\begin{equation}
  dL^2 = \frac14\Big(F_{tt}\,dt^2 + 2\,\bm f^\top d\bm\lambda\,dt
  + d\bm\lambda^\top F_{\lambda\lambda}\,d\bm\lambda\Big).
  \label{eq:SM-quad-form}
\end{equation}

For fixed $dt$, the physically relevant distance along $t$ modulo nuisance
reparametrizations is obtained by minimizing \eqref{eq:SM-quad-form} over
$d\bm\lambda\in\mathbb{R}^{m-1}$. The normal equations are
\begin{equation}
  \frac{\partial}{\partial(d\bm\lambda)} (dL^2)
  = \frac12\big(\bm f\,dt + F_{\lambda\lambda} d\bm\lambda\big) = 0,
\end{equation}
which gives
\begin{equation}
  F_{\lambda\lambda}\,d\bm\lambda_* = -\,\bm f\,dt,
  \qquad
  d\bm\lambda_* = - F_{\lambda\lambda}^{-1}\bm f\,dt.
\end{equation}
Here $F_{\lambda\lambda}^{-1}$ denotes the inverse of $F_{\lambda\lambda}$ on
its range; when $F_{\lambda\lambda}$ is singular, it is understood as the
Moore--Penrose pseudoinverse \cite{benisrael2003,HornJohnson2013}.

Substituting $d\bm\lambda_*$ back into \eqref{eq:SM-quad-form} yields the
minimal squared distance:
\begin{align}
  dL_{\min}^2
  &= \frac14\Big(
     F_{tt} - 2\,\bm f^\top F_{\lambda\lambda}^{-1}\bm f
     + \bm f^\top F_{\lambda\lambda}^{-1}
       F_{\lambda\lambda}
       F_{\lambda\lambda}^{-1}\bm f\Big)\,dt^2
   \nonumber\\
  &= \frac14\Big(F_{tt} - \bm f^\top F_{\lambda\lambda}^{-1}\bm f\Big)\,dt^2.
\end{align}
This identifies the {effective} (profiled) Fisher information for time as
the Schur complement of the nuisance block:
\begin{equation}
  F_{\rm eff}
  := F_{tt} - \bm f^\top F_{\lambda\lambda}^{-1}\bm f,
  \qquad
  dL_{\min}^2 = \frac14 F_{\rm eff}\,dt^2.
  \label{eq:SM-F-eff}
\end{equation}
Since the full QFIM $F$ is positive semidefinite, its Schur complements are
positive semidefinite as well \cite{HornJohnson2013,Zhang2005}, so
$F_{\rm eff}\ge 0$.

\paragraph*{Invariance and relation to profiled information.}
Consider a smooth reparametrization of the nuisance manifold
$\tilde{\bm\lambda}=\phi(\bm\lambda)$ with Jacobian $J$. The QFIM transforms as
$F' = J^{-\top}FJ^{-1}$, whence
\begin{align}
  F'_{\lambda\lambda}
  &= J^\top F_{\lambda\lambda} J, \\
  \bm f' &= J^\top \bm f.
\end{align}
It follows that
\begin{align}
  \bm f'^{\!\top}(F'_{\lambda\lambda})^{-1}\bm f'
  &= \bm f^\top J (J^\top F_{\lambda\lambda} J)^{-1}J^\top \bm f \\
  &= \bm f^\top F_{\lambda\lambda}^{-1}\bm f,
\end{align}
and hence $F_{\rm eff}$ is a scalar under nuisance reparametrizations.

If the full QFIM $F$ is invertible, one may also use the block inversion
formula \cite{HornJohnson2013}:
\begin{equation}
  F_{\rm eff}
  = \big( (F^{-1})_{tt} \big)^{-1},
  \label{eq:SM-profiled}
\end{equation}
which is the usual profiled Fisher information for $t$
\cite{AmariNagaoka2000InformationGeometry}.  Thus $F_{\rm eff}$ coincides with the information for
$t$ after profiling (maximizing the likelihood) over the nuisance coordinates.

\subsection{Embedding GKSL dynamics: sensitivities and QFIM blocks}
\label{subsec:SM-GKSL-sens}

We now bring the GKSL dynamics \eqref{eq:SM-GKSL} inside the geometric
construction. Treat $(t,\bm\lambda)$ as independent coordinates, and write
\begin{equation}
  d\rho
  = \partial_t\rho\,dt + \sum_\alpha \partial_{\lambda^\alpha}\rho\,d\lambda^\alpha
  \equiv \dot\rho\,dt + \sum_\alpha \rho'_\alpha\,d\lambda^\alpha,
  \label{eq:SM-differential}
\end{equation}
where $\dot\rho := \mathcal{L}_{\bm\lambda}[\rho]$ is determined by
\eqref{eq:SM-GKSL} and
\begin{equation}
  \rho'_\alpha := \partial_{\lambda^\alpha}\rho
\end{equation}
is the sensitivity of the state to a change in the nuisance parameter
$\lambda^\alpha$.

Differentiating the GKSL equation \eqref{eq:SM-GKSL} with respect to
$\lambda^\alpha$ and commuting derivatives (under the assumed smoothness) gives
\begin{equation}
  \partial_t \rho'_\alpha(t)
  = (\partial_{\lambda^\alpha}\mathcal{L}_{\bm\lambda})[\rho(t)]
    + \mathcal{L}_{\bm\lambda}[\rho'_\alpha(t)].
  \label{eq:SM-sens-ODE}
\end{equation}
This is a linear, inhomogeneous ODE on the operator space. Its solution is
given by the Duhamel formula \cite{Pazy1983,EngelNagel2000Semigroups}:
\begin{equation}
  \rho'_\alpha(t)
  = e^{t\mathcal{L}_{\bm\lambda}} \rho'_\alpha(0)
  + \int_0^t e^{(t-s)\mathcal{L}_{\bm\lambda}}\,
    (\partial_{\lambda^\alpha}\mathcal{L}_{\bm\lambda})[\rho(s,\bm\lambda)]\,ds.
  \label{eq:SM-sens-Duhamel}
\end{equation}
Trace preservation of $\mathcal{L}_{\bm\lambda}$ and of
$\partial_{\lambda^\alpha}\mathcal{L}_{\bm\lambda}$ implies that
$\Tr\,\rho'_\alpha(t)=\Tr\,\rho'_\alpha(0)$; taking
$\Tr\,\rho'_\alpha(0)=0$ ensures that $\rho'_\alpha(t)$ remains a valid
tangent (trace-zero) operator for all $t$
\cite{BreuerPetruccione2002,Pazy1983}.

Using the QFIM expression \eqref{eq:SM-QFIM-J} with
$\partial_t\rho=\mathcal{L}_{\bm\lambda}[\rho]$ and
$\partial_{\lambda^\alpha}\rho=\rho'_\alpha$, we obtain the matrix blocks along
a dynamical trajectory:
\begin{align}
  F_{tt}(t)
  &= 2\,\Tr\!\big[(\mathcal{L}_{\bm\lambda}\rho)\,
                   J_\rho^{-1}(\mathcal{L}_{\bm\lambda}\rho)\big],
  \label{eq:SM-Ftt}\\
  F_{t\alpha}(t)
  &= 2\,\Tr\!\big[(\mathcal{L}_{\bm\lambda}\rho)\,
                   J_\rho^{-1}(\rho'_\alpha)\big],
  \label{eq:SM-Ftalpha}\\
  F_{\alpha\beta}(t)
  &= 2\,\Tr\!\big[\rho'_\alpha\,J_\rho^{-1}(\rho'_\beta)\big].
  \label{eq:SM-Flambda}
\end{align}
Substituting these blocks into the Schur complement
\eqref{eq:SM-F-eff} yields the effective information $F_{\rm eff}(t)$
along the laboratory trajectory. This is the quantity that enters the
projected quantum speed limit.
{

\subsection{Quotient metric and local-to-global projected QSL}
\label{subsec:SM-quotient-local-global}

We now make explicit the quotient-geometric meaning of the profiled quantity
$F_{\rm eff}$ and its relation to the global projected quantum speed limit.

At each state $\rho$, the nuisance coordinates $\lambda^\alpha$ generate tangent
vectors $\partial_{\lambda^\alpha}\rho$. Their span defines the vertical
nuisance subspace
\begin{equation}
  \mathcal{V}_{\rho}
  :=
  \mathrm{span}\!\left\{
    \partial_{\lambda^\alpha}\rho
  \right\}_{\alpha}.
\end{equation}
On a regular patch where this nuisance distribution has constant rank, we define
the induced quadratic form on a quotient tangent class $[X]$ by minimizing the
ambient Bures metric over vertical shifts:
\begin{equation}
  g_{\rm quo}([X],[X])
  :=
  \inf_{V\in\mathcal{V}_{\rho}}
  g_{\rho}(X+V,X+V).
  \label{eq:SM-quotient-metric-def}
\end{equation}

Take now $X=\partial_t\rho$ and write
\begin{equation}
  V=\xi^\alpha\,\partial_{\lambda^\alpha}\rho .
\end{equation}
Using the block form of the QFIM, the corresponding Bures quadratic form is
\begin{equation}
  4\,g_{\rho}(X+V,X+V)
  =
  F_{tt}
  +2\,\xi^\alpha F_{t\alpha}
  +\xi^\alpha\xi^\beta F_{\alpha\beta}.
  \label{eq:SM-quotient-local-quad}
\end{equation}
Minimizing \eqref{eq:SM-quotient-local-quad} over $\xi^\alpha$ gives the normal
equations
\begin{equation}
  F_{\alpha\beta}\,\xi_*^\beta = -\,F_{t\alpha},
\end{equation}
and therefore
\begin{equation}
  4\,g_{\rm quo}\!\big([\partial_t\rho],[\partial_t\rho]\big)
  =
  F_{tt}-\bm f^\top F_{\lambda\lambda}^{-1}\bm f
  =
  F_{\rm eff}.
  \label{eq:SM-quotient-feff}
\end{equation}
Thus the Schur complement $F_{\rm eff}$ is precisely the induced quotient metric
coefficient along the time direction.

When $F_{\lambda\lambda}$ is singular, the inverse in
\eqref{eq:SM-quotient-feff} is understood in the Moore--Penrose sense on the
estimable nuisance subspace, exactly as in \eqref{eq:SM-F-eff}. Because the
full QFIM block matrix is positive semidefinite, one has
$\bm f \in \operatorname{Ran}(F_{\lambda\lambda})$, equivalently
$\bm f \perp \ker(F_{\lambda\lambda})$. Hence the Moore--Penrose
pseudoinverse yields the correct profiled minimum on the estimable nuisance
subspace. Likewise, for rank-deficient states all constructions are understood
on $\operatorname{supp}(\rho)$, as discussed in Sec.~\ref{subsec:SM-rank}.

In the SLD language, the same result can be expressed through orthogonal
projection of the time score away from the nuisance-score span. If
$\widetilde L_t$ denotes the orthogonal projection of $L_t$ onto the complement
of $\mathrm{span}\{L_\alpha\}$ with respect to the inner product
$\langle\!\langle A,B\rangle\!\rangle_\rho
:= \tfrac12\Tr[\rho\{A,B\}]$, then, when the nuisance-score Gram matrix is
singular, this projection is understood on the estimable nuisance-score
subspace, with coefficients determined by the Moore--Penrose pseudoinverse of
$F_{\lambda\lambda}$,
\begin{equation}
  \langle\!\langle \widetilde L_t,\widetilde L_t\rangle\!\rangle_\rho
  = F_{\rm eff}.
\end{equation}
Hence the profiled Fisher information and the quotient-metric coefficient
coincide.

Once this local quotient metric is fixed, the instantaneous projected speed is
the norm of the tangent to the realized evolution in the quotient geometry:
\begin{equation}
  v_{\rm quo}(t)
  =
  \sqrt{
    g_{\rm quo}\!\big([\dot\rho(t)],[\dot\rho(t)]\big)
  }
  =
  \frac12\sqrt{F_{\rm eff}(t)}.
  \label{eq:SM-quotient-speed}
\end{equation}
Therefore, along the realized trajectory $t\mapsto \rho(t)$, the quotient path
length is
\begin{equation}
  \ell_{\rm quo}[\rho(\cdot)]
  =
  \int_0^\tau \frac12\sqrt{F_{\rm eff}(t)}\,dt.
  \label{eq:SM-quotient-length}
\end{equation}
By definition of geodesic distance in the quotient geometry,
\begin{equation}
  L_{\rm quo}\big([\rho_0],[\rho_\tau]\big)
  \le
  \ell_{\rm quo}[\rho(\cdot)].
  \label{eq:SM-quotient-distance}
\end{equation}
Hence
\begin{equation}
  L_{\rm quo}\big([\rho_0],[\rho_\tau]\big)
  \le
  \int_0^\tau \frac12\sqrt{F_{\rm eff}(t)}\,dt.
  \label{eq:SM-quotient-qsl}
\end{equation}

This shows that no further optimization of $v_{\rm quo}$ along the realized path
is needed. The only additional global minimization is the one already contained
in the definition of the quotient geodesic distance $L_{\rm quo}$ itself, namely
as the infimum of quotient-path lengths over all curves connecting the endpoint
equivalence classes.}
\subsection{Projected speed and quotient quantum speed limit}
\label{subsec:SM-QSL}

As established in Sec.~\ref{subsec:SM-quotient-local-global}, the instantaneous
projected (quotient) speed is
\begin{equation}
  v_{\rm quo}(t)
  := \frac{dL_{\rm quo}}{dt}
  = \frac12\sqrt{F_{\rm eff}(t)}.
  \label{eq:SM-speed-proj}
\end{equation}
and the corresponding integral quotient-QSL is
\begin{equation}
  L_{\rm quo}\big([\rho_0],[\rho_\tau]\big)
  \;\le\; \int_0^\tau \frac12\sqrt{F_{\rm eff}(t)}\,dt.
  \label{eq:SM-proj-QSL}
\end{equation}
Rearranging \eqref{eq:SM-proj-QSL} gives the time-to-target form used in the main text,
\begin{equation}
  \tau \;\ge\;
  \frac{L_{\rm quo}\big([\rho_0],[\rho_\tau]\big)}
       {\displaystyle \bar v_{\rm quo}},
  \qquad
  \bar v_{\rm quo}
  := \frac1\tau \int_0^\tau \frac12\sqrt{F_{\rm eff}(t)}\,dt,
  \label{eq:SM-time-target}
\end{equation}
where $\bar v_{\rm quo}$ is the time-averaged projected speed. This is the form
used in the main text.

\begin{remark}[Physical vs projected QSLs]
The {physical} QSL (without profiling) pairs the physical Bures distance
$L(\rho_0,\rho_\tau)$ with the speed $\tfrac12\sqrt{F_{tt}}$, where $F_{tt}$ is
the $tt$-element of the full QFIM. The {projected} QSL pairs the quotient
distance $L_{\rm quo}$ with the projected speed
$\tfrac12\sqrt{F_{\rm eff}}$. Since $F_{\rm eff}\le F_{tt}$ and
$L_{\rm quo}\le L(\rho_0,\rho_\tau)$ (projection never increases distances),
one must not mix distances and speeds from different manifolds, as this can
invalidate the bound.
\end{remark}

\subsection{Rank-deficient states and singular nuisance blocks}
\label{subsec:SM-rank}

The derivations above assumed full rank for $\rho$ and invertibility of
$F_{\lambda\lambda}$. We now state the precise conditions under which the
construction extends to rank-deficient states and singular nuisance blocks.

\begin{lemma}[Support restriction and continuous extension]
\label{lem:SM-rank-def}
Let $\rho$ be a density matrix (possibly rank-deficient), and let
$\mathrm{supp}(\rho)$ denote its support. Restrict all constructions to
$\mathrm{supp}(\rho)$, i.e., regard $\rho$ as a strictly positive operator on
$\mathrm{supp}(\rho)$ and define $J_\rho$ and $J_\rho^{-1}$ accordingly. Then
the formulas \eqref{eq:SM-g-explicit}--\eqref{eq:SM-QFIM-J},
\eqref{eq:SM-QFIM-block}--\eqref{eq:SM-profiled},
\eqref{eq:SM-sens-ODE}--\eqref{eq:SM-Flambda}, and
\eqref{eq:SM-speed-proj}--\eqref{eq:SM-time-target} all hold with this
understanding. Moreover, continuity of the fidelity in the trace norm implies
that the resulting Bures metric is the continuous extension from
$\mathrm{supp}(\rho)$ to $\mathcal{H}$.
\end{lemma}

A detailed discussion of these support issues for the Bures metric and the
SLD--QFIM can be found in Ref.~\cite{Safranek2018}.

\begin{remark}[Singular $F_{\lambda\lambda}$]
When the nuisance block $F_{\lambda\lambda}$ is singular, the Schur complement
\eqref{eq:SM-F-eff} is defined using the Moore--Penrose pseudoinverse
$F_{\lambda\lambda}^{+}$. This yields the unique minimizer on the estimable
nuisance subspace and preserves invariance under reparametrizations of the
nuisance manifold \cite{benisrael2003,Zhang2005}.
\end{remark}

\refstepcounter{section}
\setcounter{equation}{0}
\setcounter{figure}{0}
\setcounter{table}{0}
\section*{Appendix \thesection: Closed Jaynes--Cummings sensor --- unitary projected QSL}
\label{app:JC-unitary}

In this section we specialize the general formalism of Appendix~\ref{app:formalism} to a closed
Jaynes--Cummings (JC) sensor driven by a static magnetic field. We (i) reduce
the model to a two-level single-excitation manifold; (ii) compute the local
generators for laboratory time $t$ and field $B$; (iii) evaluate the
two-parameter QFIM $\{F_{tt},F_{tB},F_{BB}\}$ exactly; (iv) form the
Schur-complement effective information $F_{\mathrm{eff}}(t)$; and (v) relate
these quantities to the fidelity and the projected (quotient) quantum speed
limit. Finally, we derive a short-time expansion that yields explicit detuning
tolerances for retaining a given fraction of the physical speed.

\subsection{Model and reduction to the single-excitation manifold}
\label{subsec:SM2-model}

We consider a single-mode cavity of frequency $\omega_c$ coupled to a
two-level system (atom or qubit) of bare transition frequency $\omega_q$ and a
magnetic-field-dependent Zeeman shift $\gamma B$. In the rotating-wave
approximation, the laboratory-frame Hamiltonian reads
\begin{equation}
  H(B) = \omega_c\,a^\dagger a
       + \frac{\omega_q}{2}\sigma_z
       + g\,(a\sigma_+ + a^\dagger\sigma_-)
       + \frac{\gamma B}{2}\,\sigma_z,
  \label{eq:SM2-Hfull}
\end{equation}
with $\hbar=1$ and the usual Pauli and ladder operators
$\sigma_z = \ket{e}\bra{e}-\ket{g}\bra{g}$,
$\sigma_+ = \ket{e}\bra{g}$, $\sigma_-=\ket{g}\bra{e}$.

We prepare the initial state
\begin{equation}
  \ket{\psi(0)} = \ket{e,0},
\end{equation}
and assume that throughout the relevant evolution time the dynamics remains in
the single-excitation manifold spanned by $\{\ket{e,0},\ket{g,1}\}$. In this
manifold the total excitation number
\begin{equation}
  N = a^\dagger a + \sigma_+\sigma_-
\end{equation}
is conserved and equals $1$.

In the ordered basis $\{\ket{e,0},\ket{g,1}\}$ we have the matrix
representations
\begin{equation}
\begin{aligned}
 a^\dagger a &\mapsto \begin{pmatrix}0&0\\0&1\end{pmatrix},\qquad
 \sigma_z \mapsto \begin{pmatrix}1&0\\0&-1\end{pmatrix},\\
 a\sigma_++a^\dagger\sigma_- &\mapsto \begin{pmatrix}0&1\\1&0\end{pmatrix}.
\end{aligned}
\end{equation}
Define the {detuning}
\begin{equation}
  \Delta(B) := \omega_q + \gamma B - \omega_c,
  \label{eq:SM2-Delta}
\end{equation}
and subtract the scalar term $(\omega_c/2)\,\mathbb{I}$ from $H(B)$: this does
not affect any QFI element or variance, since adding $c\,\mathbb{I}$ to a
Hamiltonian leaves commutators and variances invariant. We thus work with the
traceless effective two-level Hamiltonian
\begin{equation}
  H'(B) = g\,\tau_x + \frac{\Delta(B)}{2}\,\tau_z,
  \label{eq:SM2-Hprime}
\end{equation}
where $\bm\tau=(\tau_x,\tau_y,\tau_z)$ are Pauli matrices acting on
$\{\ket{e,0},\ket{g,1}\}$. Introducing the generalized Rabi frequency
\begin{equation}
  \Omega(B) := \sqrt{\Delta(B)^2 + 4g^2},
  \label{eq:SM2-Omega}
\end{equation}
we can write
\begin{equation}
  H'(B) = \frac{\Omega(B)}{2}\,\bm n(B)\!\cdot\!\bm\tau,\qquad
  \bm n(B) = \frac{1}{\Omega(B)}\,(2g,\,0,\,\Delta(B)).
  \label{eq:SM2-Hprime-n}
\end{equation}
The unitary time evolution in this manifold is
\begin{equation}
  U(t;B) = e^{-i H'(B) t}.
  \label{eq:SM2-U}
\end{equation}
The initial state $\ket{\psi(0)}=\ket{e,0}$ is the $+1$ eigenstate of $\tau_z$,
which we denote by $\ket{+z}$ in what follows.

\subsection{Local generators and QFI for $(t,B)$}
\label{subsec:SM2-local-generators}

For a pure state family $\ket{\psi(x)} = U(x)\ket{\psi_0}$ with parameters
$x^\mu$, the SLD QFIM is given by the covariance of the local generators
$G_\mu$ defined by
\begin{equation}
  i\,\partial_\mu U(x) = G_\mu(x)\,U(x).
\end{equation}
Explicitly \cite{braunstein1994},
\begin{equation}
  F_{\mu\nu}
  = 4\,\mathrm{Cov}_{\psi_0}(G_\mu,G_\nu)
  := 4\,\Re\Bigl(\braket{\psi_0|G_\mu G_\nu|\psi_0}
    - \braket{\psi_0|G_\mu|\psi_0}\braket{\psi_0|G_\nu|\psi_0}\Bigr),
  \label{eq:SM2-QFI-pure}
\end{equation}
where all expectations are taken in the {initial} state
$\ket{\psi_0}=\ket{+z}$.

In our case the parameters are $(t,B)$ and
\begin{equation}
  U(t,B) = e^{-i H'(B)t}.
\end{equation}
The time generator is simply the Hamiltonian,
\begin{equation}
  G_t = H'(B).
  \label{eq:SM2-Gt}
\end{equation}
The field generator $G_B(t)$ is obtained by differentiating $U(t;B)$ with
respect to $B$:
\begin{align}
  \partial_B U(t;B)
  &= -i \int_0^t U(t;B) U^\dagger(s;B)\,\partial_B H'(B)\,U(s;B)\,ds \nonumber\\
  &= -i\,U(t;B) \int_0^t U^\dagger(s;B)\,\partial_B H'(B)\,U(s;B)\,ds.
\end{align}
Thus the local generator for $B$ is
\begin{equation}
  G_B(t) = \int_0^t U^\dagger(s;B)\,\partial_B H'(B)\,U(s;B)\,ds.
  \label{eq:SM2-GB-def}
\end{equation}
From \eqref{eq:SM2-Hprime}, the $B$-dependence enters only through the
detuning, so that
\begin{equation}
  \partial_B H'(B) = \frac{\gamma}{2}\,\tau_z.
  \label{eq:SM2-Hprime-deriv}
\end{equation}
Substituting \eqref{eq:SM2-Hprime-deriv} into \eqref{eq:SM2-GB-def} gives
\begin{equation}
  G_B(t) = \frac{\gamma}{2}\int_0^t \tau_z(s)\,ds,
  \qquad
  \tau_z(s) := U^\dagger(s;B)\,\tau_z\,U(s;B).
  \label{eq:SM2-GB-rot}
\end{equation}
We therefore need the exact Heisenberg evolution of $\tau_z$ under $H'$.

The relevant moments in the initial state $\ket{+z}$ are
\begin{equation}
\begin{aligned}
  \langle\tau_z\rangle &= 1, &
  \langle\tau_x\rangle &= \langle\tau_y\rangle = 0,\\
  \langle\tau_i\tau_j\rangle &= \delta_{ij}, &
  \Re\langle\tau_i\tau_j\rangle &= 0 \quad (i\neq j),
\end{aligned}
\label{eq:SM2-zstate-moments}
\end{equation}
which we will use repeatedly below.

\subsection{Heisenberg evolution of $\tau_z$ and closed-form $G_B(t)$}
\label{subsec:SM2-Heisenberg}

The effective Hamiltonian \eqref{eq:SM2-Hprime-n} shows that the dynamics is a
rotation on the Bloch sphere about the axis $\bm n(B)$ with angular frequency
$\Omega(B)$. For any vector $\bm v\in\mathbb{R}^3$,
\begin{equation}
  e^{+i(\Omega/2)\bm n\cdot\bm\tau\,s}\,(\bm v\!\cdot\!\bm\tau)\,
  e^{-i(\Omega/2)\bm n\cdot\bm\tau\,s}
  = \bigl(R_{\bm n}(\Omega s)\bm v\bigr)\!\cdot\!\bm\tau,
  \label{eq:SM2-rot-general}
\end{equation}
where $R_{\bm n}(\theta)$ is the rotation matrix about $\bm n$ by angle
$\theta$. Applying \eqref{eq:SM2-rot-general} to $\bm v=\hat{\bm z}$ yields
\begin{equation}
  \tau_z(s) = v_x(s)\,\tau_x + v_y(s)\,\tau_y + v_z(s)\,\tau_z,
  \label{eq:SM2-tauz-decomp}
\end{equation}
with coefficients determined by Rodrigues' formula
\begin{equation}
  R_{\bm n}(\theta)\bm v = \bm n(\bm n\!\cdot\!\bm v)
  + (\bm v - \bm n\,\bm n\!\cdot\!\bm v)\cos\theta
  + (\bm n\times\bm v)\sin\theta.
\end{equation}
Using $\bm n=(2g/\Omega,\,0,\,\Delta/\Omega)$ and $\bm v=\hat{\bm z}$,
a straightforward calculation gives
\begin{align}
  v_x(s) &= \frac{2g\Delta}{\Omega^2}\Bigl(1-\cos\Omega s\Bigr), \nonumber\\
  v_y(s) &= -\,\frac{2g}{\Omega}\sin\Omega s, \nonumber\\
  v_z(s) &= \frac{\Delta^2}{\Omega^2}
          + \frac{4g^2}{\Omega^2}\cos\Omega s,
  \label{eq:SM2-vxyz}
\end{align}
where we suppress the explicit $B$-dependence of $\Delta$ and $\Omega$ for
readability.

Inserting \eqref{eq:SM2-tauz-decomp} and \eqref{eq:SM2-vxyz} into
\eqref{eq:SM2-GB-rot}, we obtain
\begin{equation}
  G_B(t) = \frac{\gamma}{2}\Bigl(A_x(t)\,\tau_x
                               + A_y(t)\,\tau_y
                               + A_z(t)\,\tau_z\Bigr),
  \label{eq:SM2-GB-ABC}
\end{equation}
with the time-dependent coefficients
\begin{align}
  A_x(t) &= \int_0^t v_x(s)\,ds
         = \frac{2g\Delta}{\Omega^2}
            \Bigl(t-\frac{\sin\Omega t}{\Omega}\Bigr),
  \label{eq:SM2-Ax}\\
  A_y(t) &= \int_0^t v_y(s)\,ds
         = -\,\frac{2g}{\Omega^2}\Bigl(1-\cos\Omega t\Bigr),
  \label{eq:SM2-Ay}\\
  A_z(t) &= \int_0^t v_z(s)\,ds
         = \frac{\Delta^2}{\Omega^2}\,t
           + \frac{4g^2}{\Omega^3}\sin\Omega t.
  \label{eq:SM2-Az}
\end{align}
No approximations have been made: the only step between
\eqref{eq:SM2-tauz-decomp} and \eqref{eq:SM2-Ax}--\eqref{eq:SM2-Az} is the
evaluation of elementary trigonometric integrals.

\subsection{Exact QFIM entries $F_{tt}$, $F_{BB}$, and $F_{tB}$}
\label{subsec:SM2-QFIM-entries}

We now evaluate the QFIM elements for the parameter pair $(t,B)$ using
\eqref{eq:SM2-QFI-pure} and the local generators
\eqref{eq:SM2-Gt} and \eqref{eq:SM2-GB-ABC}, with expectations taken in the
initial state $\ket{+z}$.

\paragraph*{(i) The time-time element $F_{tt}$.}
From \eqref{eq:SM2-Hprime} and \eqref{eq:SM2-zstate-moments},
\begin{align}
  \mathrm{Var}(H')
  &= \left\langle\bigl(g\tau_x+\tfrac{\Delta}{2}\tau_z\bigr)^2\right\rangle
   - \left\langle g\tau_x+\tfrac{\Delta}{2}\tau_z\right\rangle^2 \nonumber\\
  &= g^2\langle\tau_x^2\rangle
     + \frac{\Delta^2}{4}\langle\tau_z^2\rangle
     + g\Delta\,\Re\langle\tau_x\tau_z\rangle
     - \Bigl(0 + \tfrac{\Delta}{2}\cdot1\Bigr)^2 \nonumber\\
  &= g^2 + \frac{\Delta^2}{4} - \frac{\Delta^2}{4} = g^2,
\end{align}
so that
\begin{equation}
  F_{tt} = 4\,\mathrm{Var}(H') = 4g^2.
  \label{eq:SM2-Ftt}
\end{equation}
Equivalently, one may note that $(H')^2=(\Delta^2/4+g^2)\,\mathbb{I}$, so the
variance in a pure state reduces to $g^2$.

\paragraph*{(ii) The field-field element $F_{BB}$.}
For any vector $\bm b\in\mathbb{R}^3$,
$(\bm b\!\cdot\!\bm\tau)^2 = \|\bm b\|^2\,\mathbb{I}$, as the antisymmetric
part drops out, and hence
\begin{align}
  \langle G_B^2\rangle
  &= \Bigl(\frac{\gamma}{2}\Bigr)^2\langle(A_x\tau_x+A_y\tau_y+A_z\tau_z)^2\rangle
   = \Bigl(\frac{\gamma}{2}\Bigr)^2 \bigl(A_x^2 + A_y^2 + A_z^2\bigr),\\
  \langle G_B\rangle
  &= \frac{\gamma}{2}\bigl(A_x\langle\tau_x\rangle
                          + A_y\langle\tau_y\rangle
                          + A_z\langle\tau_z\rangle\bigr)
   = \frac{\gamma}{2}A_z,
\end{align}
where we used \eqref{eq:SM2-zstate-moments}. Therefore
\begin{align}
  \mathrm{Var}(G_B)
  &= \langle G_B^2\rangle - \langle G_B\rangle^2
   = \Bigl(\frac{\gamma}{2}\Bigr)^2\bigl(A_x^2 + A_y^2\bigr),
\end{align}
and
\begin{equation}
  F_{BB}(t) = 4\,\mathrm{Var}(G_B) = \gamma^2\bigl[A_x(t)^2 + A_y(t)^2\bigr].
  \label{eq:SM2-FBB}
\end{equation}
Substituting \eqref{eq:SM2-Ax} and \eqref{eq:SM2-Ay}, we obtain the explicit
closed form
\begin{align}
  F_{BB}(t)
  &= \frac{4g^2\gamma^2}{\Omega^4}
     \left[\Delta^2\!\left(t-\frac{\sin\Omega t}{\Omega}\right)^{\!2}
           + \Bigl(1-\cos\Omega t\Bigr)^{\!2}\right] \nonumber\\
  &= \frac{4g^2\gamma^2}{\Omega^4}
     \left[\Delta^2\!\left(t-\frac{\sin\Omega t}{\Omega}\right)^{\!2}
           + 4\sin^4\!\Bigl(\frac{\Omega t}{2}\Bigr)\right].
  \label{eq:SM2-FBB-explicit}
\end{align}

\paragraph*{(iii) The mixed element $F_{tB}$.}
By \eqref{eq:SM2-QFI-pure}, the mixed QFI element is
\begin{equation}
  F_{tB} = 4\,\mathrm{Cov}(H',G_B)
  = 4\,\Re\bigl(\langle H'G_B\rangle
                - \langle H'\rangle\langle G_B\rangle\bigr).
\end{equation}
Using $H' = g\tau_x + (\Delta/2)\tau_z$ and $G_B$ from
\eqref{eq:SM2-GB-ABC}, with expectations in $\ket{+z}$,
\begin{align}
  \mathrm{Cov}(H',G_B)
  &= \frac{\gamma}{2}\,\Re\left\langle
        \bigl(g\tau_x+\tfrac{\Delta}{2}\tau_z\bigr)
        \bigl(A_x\tau_x+A_y\tau_y+A_z\tau_z\bigr)
     \right\rangle
     - \frac{\gamma}{2}\Bigl(\tfrac{\Delta}{2}\Bigr)A_z \nonumber\\
  &= \frac{\gamma}{2}\,\Re\bigl(
     gA_x\langle\tau_x\tau_x\rangle
     + gA_y\langle\tau_x\tau_y\rangle
     + \tfrac{\Delta}{2}A_z\langle\tau_z\tau_z\rangle\bigr)
     - \frac{\gamma}{2}\,\frac{\Delta}{2}A_z.
\end{align}
Using \eqref{eq:SM2-zstate-moments},
$\langle\tau_x\tau_x\rangle=\langle\tau_z\tau_z\rangle=1$ and
$\Re\langle\tau_x\tau_y\rangle=0$, we get
\begin{equation}
  \mathrm{Cov}(H',G_B) = \frac{\gamma}{2}\,gA_x.
\end{equation}
Therefore
\begin{equation}
  F_{tB}(t)
  = 4\,\mathrm{Cov}(H',G_B)
  = 2\gamma g\,A_x(t)
  = \frac{4g^2\gamma\,\Delta}{\Omega^2}
    \left(t-\frac{\sin\Omega t}{\Omega}\right).
  \label{eq:SM2-FtB}
\end{equation}

Collecting the results \eqref{eq:SM2-Ftt}, \eqref{eq:SM2-FBB-explicit}, and
\eqref{eq:SM2-FtB}, the full $2\times2$ QFIM for $(t,B)$ is
\begin{align}
  F_{tt} &= 4g^2, \label{eq:SM2-Ftt-collect}\\
  F_{BB}(t) &=
  \frac{4g^2\gamma^2}{\Omega^4}\left[\Delta^2\left(t-\frac{\sin \Omega t}{\Omega}\right)^{\!2}
  + 4\sin^4\!\left(\frac{\Omega t}{2}\right)\right],
  \label{eq:SM2-FBB-collect}\\
  F_{tB}(t) &=
  \frac{4g^2\gamma\,\Delta}{\Omega^2}\left(t-\frac{\sin \Omega t}{\Omega}\right),
  \label{eq:SM2-FtB-collect}
\end{align}
with $\Omega=\sqrt{\Delta^2+4g^2}$ and $\Delta=\omega_q+\gamma B-\omega_c$.

\subsection{Schur-complement effective information $F_{\mathrm{eff}}(t)$}
\label{subsec:SM2-Feff}

We now treat $t$ as the parameter of interest and $B$ as a nuisance, and apply
the Schur complement of Appendix~\ref{app:formalism} to obtain the effective information for time.
With a single nuisance parameter, the Schur complement reduces to
\begin{equation}
  F_{\mathrm{eff}}(t)
  = F_{tt} - \frac{F_{tB}(t)^2}{F_{BB}(t)},
  \label{eq:SM2-Feff-def}
\end{equation}
understanding $F_{BB}^{-1}$ as the Moore--Penrose pseudoinverse when
$F_{BB}=0$ (in which case necessarily $F_{tB}=0$ as well).

Substituting \eqref{eq:SM2-Ftt-collect}--\eqref{eq:SM2-FtB-collect} and using
$A_x$ and $A_y$ defined in \eqref{eq:SM2-Ax}--\eqref{eq:SM2-Ay}, it is
convenient to write $F_{BB}=\gamma^2(A_x^2+A_y^2)$ and
$F_{tB}=2\gamma gA_x$. Then
\begin{align}
  F_{\mathrm{eff}}(t)
  &= 4g^2 - \frac{4\gamma^2 g^2 A_x(t)^2}{\gamma^2\bigl(A_x(t)^2+A_y(t)^2\bigr)} \nonumber\\
  &= 4g^2\,\frac{A_y(t)^2}{A_x(t)^2+A_y(t)^2}.
  \label{eq:SM2-Feff-compact}
\end{align}
Using the explicit expressions for $A_x$ and $A_y$,
\begin{align}
  A_x(t)
  &= \frac{2g\Delta}{\Omega^2}\left(t-\frac{\sin\Omega t}{\Omega}\right),\\
  A_y(t)
  &= -\,\frac{2g}{\Omega^2}\left(1-\cos\Omega t\right),
\end{align}
we can write $F_{\mathrm{eff}}(t)$ in closed form:
\begin{align}
  F_{\mathrm{eff}}(t)
  &= 4g^2\,
     \frac{\bigl(1-\cos(\Omega t)\bigr)^2}
          {\Delta^2\left(t-\tfrac{\sin(\Omega t)}{\Omega}\right)^2
           + \bigl(1-\cos(\Omega t)\bigr)^2} \nonumber\\
  &= 4g^2\,
     \frac{4\sin^{4}\!\bigl(\tfrac{\Omega t}{2}\bigr)}
          {\displaystyle
           \Delta^2\left(t-\frac{\sin(\Omega t)}{\Omega}\right)^{\!2}
           + 4\sin^{4}\!\bigl(\tfrac{\Omega t}{2}\bigr)}.
  \label{eq:SM2-Feff-explicit}
\end{align}
By construction, $0\le F_{\mathrm{eff}}(t)\le F_{tt}=4g^2$, and equality
$F_{\mathrm{eff}}(t)=4g^2$ holds whenever $A_x(t)=0$, i.e., when the mixed
block $F_{tB}$ vanishes.

The instantaneous projected speed is then
\begin{equation}
  v_{\mathrm{quo}}(t)
  = \frac12\sqrt{F_{\mathrm{eff}}(t)}
  = g\,\sqrt{\frac{A_y(t)^2}{A_x(t)^2+A_y(t)^2}},
  \label{eq:SM2-vquo}
\end{equation}
and the time-averaged projected speed is
\begin{equation}
  \bar v_{\mathrm{quo}}
  = \frac{1}{\tau}\int_0^\tau \frac12\sqrt{F_{\mathrm{eff}}(t)}\,dt.
  \label{eq:SM2-vquo-avg}
\end{equation}

\subsection{Fidelity, Bures angle, and quotient distance}
\label{subsec:SM2-fidelity}

To connect the effective information to an operational distance, we compute the
fidelity between the initial state $\ket{\psi(0)}=\ket{+z}$ and the evolved
state $\ket{\psi(t,B)}=U(t;B)\ket{+z}$. Using
$H'=\frac{\Omega}{2}\bm n\!\cdot\!\bm\tau$ and \eqref{eq:SM2-Hprime-n}, the
propagator is
\begin{equation}
  U(t;B)
  = e^{-i\frac{\Omega t}{2}\,\bm n\cdot\bm\tau}
  = \cos\!\frac{\Omega t}{2}\,\mathbb{I}
    - i\sin\!\frac{\Omega t}{2}\,\bm n\!\cdot\!\bm\tau,
  \label{eq:SM2-U-explicit}
\end{equation}
with $\bm n=(2g,0,\Delta)/\Omega$. The overlap with $\ket{+z}$ is
\begin{align}
  \braket{\psi(0)|\psi(t,B)}
  &= \braket{+z|U(t;B)|+z} \nonumber\\
  &= \cos\!\frac{\Omega t}{2}
   - i\,\frac{\Delta}{\Omega}\,\sin\!\frac{\Omega t}{2},
\end{align}
where we used $\braket{+z|\bm n\!\cdot\!\bm\tau|+z}=n_z=\Delta/\Omega$. The
fidelity is therefore
\begin{align}
  F(t,B)
  &= \bigl|\braket{\psi(0)|\psi(t,B)}\bigr|^2 \nonumber\\
  &= \cos^2\!\frac{\Omega t}{2} + \Bigl(\frac{\Delta}{\Omega}\Bigr)^2
      \sin^2\!\frac{\Omega t}{2} \nonumber\\
  &= 1 - \frac{4g^2}{\Omega^2}\sin^2\!\Bigl(\frac{\Omega t}{2}\Bigr).
  \label{eq:SM2-fidelity}
\end{align}
This is the closed-system survival fidelity for the JC sensor at detuning
$\Delta(B)$.

When $B$ is treated as a nuisance, the relevant distance on the quotient
manifold is obtained by optimizing the fidelity over admissible $B$ values at
fixed interrogation time $\tau$. The quotient Bures angle is
\begin{equation}
  \Theta_{\rm quo}^{\rm (uni)}(\tau)
  := \arccos\!\Big(
      \sup_{B\in\mathcal{B}}
      \sqrt{F(\tau,B)}\Big),
  \label{eq:SM2-theta-quo-uni-def}
\end{equation}
where $\mathcal{B}$ is the experimentally admissible set of fields (set by
calibration and hardware constraints). Using \eqref{eq:SM2-fidelity}, this can
be written as
\begin{align}
  \Theta_{\rm quo}^{\rm (uni)}(\tau)
  &= \arccos\!\left(
     \sqrt{\,1 - \inf_{B\in\mathcal{B}}
                 \frac{4g^2}{\Omega(B)^2}
                 \sin^2\!\Bigl(\tfrac{\Omega(B)\tau}{2}\Bigr)}\right).
  \label{eq:SM2-theta-quo-uni}
\end{align}
If some $B\in\mathcal{B}$ satisfies $\Omega(B)\tau\in 2\pi\mathbb{Z}$, the
infimum evaluates to $0$ and $\Theta_{\rm quo}^{\rm (uni)}(\tau)=0$, reflecting
the fact that the field may be chosen to bring the system back to its initial
state at time $\tau$.

The projected QSL then reads
\begin{equation}
  \tau \;\ge\;
  \frac{\Theta_{\rm quo}^{\rm (uni)}(\tau)}
       {\displaystyle \bar v_{\mathrm{quo}}},
  \label{eq:SM2-QSL-uni}
\end{equation}
with $\bar v_{\mathrm{quo}}$ given by \eqref{eq:SM2-vquo-avg}. This is the
closed-system counterpart of the general projected QSL of Appendix~\ref{app:formalism}.

\subsection{Short-time expansion and detuning tolerances}
\label{subsec:SM2-short-time}

For short evolution times where $|\Omega t|\ll 1$ and $|\Delta t|\ll 1$, the
ratio of the effective to the physical Fisher information admits a simple
approximate expression that quantifies the impact of detuning on the projected
speed. From \eqref{eq:SM2-Feff-compact},
\begin{equation}
  \frac{F_{\mathrm{eff}}(t)}{F_{tt}}
  = \frac{A_y(t)^2}{A_x(t)^2 + A_y(t)^2},
  \label{eq:SM2-Feff-ratio}
\end{equation}
with $A_x$ and $A_y$ given by \eqref{eq:SM2-Ax}--\eqref{eq:SM2-Ay}. Expanding
\begin{equation}
  \sin(\Omega t) \approx \Omega t - \frac{(\Omega t)^3}{6},\qquad
  \cos(\Omega t) \approx 1 - \frac{(\Omega t)^2}{2},
\end{equation}
we obtain to leading nontrivial orders
\begin{align}
  A_y(t)
  &= -\,\frac{2g}{\Omega^2}\bigl(1-\cos\Omega t\bigr)
   \approx -\,\frac{2g}{\Omega^2}\,\frac{(\Omega t)^2}{2}
   = -\,g\,t^2 + O(t^4),
  \label{eq:SM2-Ay-short}\\
  A_x(t)
  &= \frac{2g\Delta}{\Omega^2}
     \left(t-\frac{\sin\Omega t}{\Omega}\right)
   \approx \frac{2g\Delta}{\Omega^2}\,\frac{(\Omega t)^3}{6}
   = \frac{g\Delta}{3}\,t^3 + O(t^5).
  \label{eq:SM2-Ax-short}
\end{align}
Therefore
\begin{equation}
  \frac{A_x(t)^2}{A_y(t)^2}
  \approx \frac{\bigl((g\Delta/3)t^3\bigr)^2}{(g t^2)^2}
  = \left(\frac{\Delta t}{3}\right)^2,
\end{equation}
and substituting into \eqref{eq:SM2-Feff-ratio} gives
\begin{equation}
  \frac{F_{\mathrm{eff}}(t)}{F_{tt}}
  = \frac{1}{1 + A_x^2/A_y^2}
  \approx \frac{1}{1 + (\Delta t/3)^2}.
  \label{eq:SM2-Feff-ratio-short}
\end{equation}
Thus, at short times the projected (effective) speed satisfies
\begin{equation}
  \frac{v_{\mathrm{quo}}(t)}{\tfrac12\sqrt{F_{tt}}}
  = \sqrt{\frac{F_{\mathrm{eff}}(t)}{F_{tt}}}
  \approx \frac{1}{\sqrt{1+(\Delta t/3)^2}},
\end{equation}
showing that any nonzero detuning reduces the operational speed, with a
quadratic penalty in the product $|\Delta|t$.

To maintain the effective information at a fraction $R\in(0,1]$ of the physical
one (i.e., $F_{\mathrm{eff}}/F_{tt}\ge R$), the short-time approximation
\eqref{eq:SM2-Feff-ratio-short} implies the tolerance condition
\begin{equation}
  \frac{1}{1 + (\Delta t/3)^2} \;\ge\; R
  \quad\Longrightarrow\quad
  |\Delta|\,t < 3\sqrt{\frac{1}{R}-1}.
  \label{eq:SM2-tolerance}
\end{equation}
This sets a quantitative requirement on how close to resonance the system must
be kept for a given interrogation time $t$ in order to retain a target fraction
$R$ of the physical QSL speed.

For example, using \eqref{eq:SM2-tolerance}, one finds
\begin{center}
\begin{tabular}{c@{\quad}c}
\hline
Retention $R$ & Tolerance condition on $|\Delta|\,t$ \\
\hline\hline
$R=0.99$ & $|\Delta|\,t < 0.30$ \\
$R=0.95$ & $|\Delta|\,t < 0.69$ \\
$R=0.90$ & $|\Delta|\,t < 1.00$ \\
\hline
\end{tabular}
\end{center}
valid in the regime $|\Delta|\,t\lesssim 1$. These relations can be used to
translate metrological design targets (e.g., “retain at least $95\%$ of the
speed-limit tightness”) into concrete specifications on detuning control,
calibration effort, and maximal interrogation time for the JC sensor.

\refstepcounter{section}
\setcounter{equation}{0}
\setcounter{figure}{0}
\setcounter{table}{0}
\section*{Appendix \thesection: Open Jaynes--Cummings sensor --- dispersive lossy case}
\label{app:JC-open}

In this section, we embed the projected speed-limit formalism of Appendix~\ref{app:formalism} into a
realistic open Jaynes--Cummings (JC) sensor operating in the dispersive regime.
We (i) specify the full atom--cavity Hamiltonian with Markovian dissipation;
(ii) carry out a Schrieffer--Wolff (dispersive) transformation of both the
Hamiltonian and the Lindblad operators, making all approximations explicit;
(iii) derive an effective master equation for a truncated cavity qubit with a
field-dependent frequency and Purcell-enhanced decay rate; (iv) solve this
reduced dynamics exactly in Bloch form; (v) compute the quantum Fisher
information (QFI) for laboratory time $t$ and field $B$, together with the
mixed element $F_{tB}$; and (vi) build the Schur-complement effective
information $F_{\mathrm{eff}}(t)$ and the corresponding projected
(quotient) QSL.

Throughout this Appendix we work in units $\hbar=1$.

\subsection{Full JC model with Markovian dissipation}
\label{subsec:SM3-model}

We consider a two-level atom (or qubit) with ground and excited states
$\ket{g}$ and $\ket{e}$, coupled to a single cavity mode truncated to the
lowest two Fock states $\{\ket{0},\ket{1}\}$. The total Hilbert space is
\begin{equation}
  \mathcal{H}
  = \mathcal{H}_\text{cav}^{(2)} \otimes \mathcal{H}_\text{atom}
  = \operatorname{span}\{\ket{0},\ket{1}\}
    \otimes \operatorname{span}\{\ket{g},\ket{e}\}.
\end{equation}
On the truncated cavity space we use
\begin{equation}
  \tilde a = \ket{0}\!\bra{1},\qquad
  \tilde a^\dagger = \ket{1}\!\bra{0},\qquad
  \tilde a^\dagger\tilde a = \ket{1}\!\bra{1}.
\end{equation}
The atomic operators are the usual Pauli matrices
$\sigma_z = \ket{e}\bra{e}-\ket{g}\bra{g}$ and
$\sigma_+ = \ket{e}\bra{g}$, $\sigma_- = \ket{g}\bra{e}$.

The laboratory-frame Hamiltonian in the rotating-wave approximation is
\begin{equation}
  H =
  \omega_c\,\tilde a^\dagger \tilde a
  + \frac{\omega'_a(B)}{2} \sigma_z
  + g\bigl(\sigma_+ \tilde a + \sigma_- \tilde a^\dagger\bigr),
  \label{eq:SM3-H-lab}
\end{equation}
with cavity frequency $\omega_c$, atom--cavity coupling $g$, and a
field-dependent atomic transition frequency
\begin{equation}
  \omega'_a(B) = \omega_a + \eta B,
  \label{eq:SM3-omega-a-prime}
\end{equation}
where $\eta$ is a Zeeman coefficient (e.g.\ $\eta = 2\mu_B$ in suitable
units). The magnetic field $B$ is the sensing parameter.

Dissipation is described by a GKSL generator with three Markovian channels:
\begin{align}
  L_\kappa    &= \sqrt{\kappa}\,\tilde a
      &&\text{(cavity loss)}, \nonumber\\
  L_{\gamma_1} &= \sqrt{\gamma_1}\,\sigma_-
      &&\text{(atomic relaxation)}, \nonumber\\
  L_{\gamma_\varphi} &= \sqrt{\gamma_\varphi/2}\,\sigma_z
      &&\text{(atomic dephasing)}.
\end{align}
The full master equation is
\begin{equation}
  \dot\rho
  = -i[H,\rho]
    + \kappa\,\mathcal{D}[\tilde a]\rho
    + \gamma_1\,\mathcal{D}[\sigma_-]\rho
    + \gamma_\varphi\,\mathcal{D}[\sigma_z]\rho,
  \label{eq:SM3-master-full}
\end{equation}
where the dissipator is
\begin{equation}
  \mathcal{D}[L]\rho
  := L\rho L^\dagger - \frac12\{L^\dagger L,\rho\}.
\end{equation}
This is of the GKSL form
$\dot\rho = \mathcal{L}_B[\rho]$ used in Appendix~\ref{app:formalism}, with $B$ the nuisance parameter
and $t$ the laboratory time.

We work in the dispersive regime
\begin{equation}
  |\Delta(B)| \gg g,\qquad
  \Delta(B) := \omega_c - \omega'_a(B),
  \label{eq:SM3-dispersive-regime}
\end{equation}
and keep terms up to second order in the small parameter
$\lambda := g/\Delta(B)$ when performing the dispersive transformation.

\subsection{Dispersive (Schrieffer--Wolff) transformation}
\label{subsec:SM3-SW}

To eliminate the transverse coupling and expose the dispersive shift and
Purcell effect, we implement a Schrieffer--Wolff transformation generated by
\begin{equation}
  S = \lambda\bigl(\sigma_+ \tilde a - \sigma_- \tilde a^\dagger\bigr),
  \qquad
  \lambda = \frac{g}{\Delta(B)}.
  \label{eq:SM3-S-generator}
\end{equation}
The unitary transformation is $T = e^S$. The transformed Hamiltonian is
\begin{equation}
  H' = e^{-S} H e^{S}
  \approx H + [H,S] + \frac12[[H,S],S],
  \label{eq:SM3-Hprime-BCH}
\end{equation}
where we keep terms up to $O(\lambda^2)$.

A straightforward but standard calculation (see e.g. \cite{carmichael1993,gardiner2004})
gives the dispersive Hamiltonian
\begin{equation}
  H' = \omega_c\,\tilde a^\dagger \tilde a
     + \frac{1}{2}\bigl[\omega'_a(B) + \chi(B)\bigr]\sigma_z
     + \chi(B)\,\sigma_z \tilde a^\dagger \tilde a,
  \label{eq:SM3-Hprime-dispersive}
\end{equation}
with the AC Stark (dispersive) shift
\begin{equation}
  \chi(B) = \frac{g^2}{\Delta(B)}.
  \label{eq:SM3-chi}
\end{equation}
Equation \eqref{eq:SM3-Hprime-dispersive} removes the transverse coupling to
leading order, replacing it with a state-dependent cavity frequency and a
state-dependent atomic frequency.

\subsection{Polaron-transformed dissipators and Purcell channel}
\label{subsec:SM3-dissipators}

The dispersive transformation must also be applied to the Lindblad operators.
For any system operator $L$, we define
\begin{equation}
  L' := T^\dagger L T \approx L + [L,S] + O(\lambda^2).
\end{equation}
Retaining terms to first order in $\lambda$ gives the leading corrections
generated by the atom--cavity coupling inside the dissipators.

\paragraph*{Cavity loss.}
For $L_\kappa = \sqrt{\kappa}\,\tilde a$,
\begin{align}
  [L_\kappa,S]
  &= \bigl[\sqrt{\kappa}\,\tilde a,\,
           \lambda(\sigma_+ \tilde a - \sigma_- \tilde a^\dagger)\bigr] \nonumber\\
  &= -\,\lambda\sqrt{\kappa}\,\sigma_- [\tilde a,\tilde a^\dagger] \nonumber\\
  &= -\,\lambda\sqrt{\kappa}\,\sigma_-.
\end{align}
Thus
\begin{equation}
  L_\kappa' \approx \sqrt{\kappa}\bigl(\tilde a + \lambda\sigma_-\bigr).
  \label{eq:SM3-Lkappa-prime}
\end{equation}
Physically, cavity loss now induces an additional decay path for the atom.

\paragraph*{Atomic relaxation.}
For $L_{\gamma_1}=\sqrt{\gamma_1}\,\sigma_-$,
\begin{align}
  [L_{\gamma_1},S]
  &= \bigl[\sqrt{\gamma_1}\,\sigma_-,\,
           \lambda(\sigma_+ \tilde a - \sigma_- \tilde a^\dagger)\bigr] \nonumber\\
  &= \sqrt{\gamma_1}\lambda\,[\sigma_-,\sigma_+]\,\tilde a \nonumber\\
  &= -\,\sqrt{\gamma_1}\lambda\,\sigma_z \tilde a.
\end{align}
Hence
\begin{equation}
  L_{\gamma_1}' \approx \sqrt{\gamma_1}\bigl(\sigma_- - \lambda\sigma_z\tilde a\bigr),
  \label{eq:SM3-Lgamma1-prime}
\end{equation}
which couples the atomic relaxation channel to the cavity field.

\paragraph*{Atomic dephasing.}
For $L_{\gamma_\varphi}
=\sqrt{\gamma_\varphi/2}\,\sigma_z$,
\begin{align}
  [L_{\gamma_\varphi},S]
  &= \Bigl[\sqrt{\tfrac{\gamma_\varphi}{2}}\,\sigma_z,\,
           \lambda(\sigma_+ \tilde a - \sigma_- \tilde a^\dagger)\Bigr] \nonumber\\
  &= \sqrt{\tfrac{\gamma_\varphi}{2}}\,\lambda
     \Bigl([\sigma_z,\sigma_+]\,\tilde a - [\sigma_z,\sigma_-]\,\tilde a^\dagger\Bigr) \nonumber\\
  &= \sqrt{\tfrac{\gamma_\varphi}{2}}\,\lambda
     \Bigl(2\sigma_+\tilde a + 2\sigma_- \tilde a^\dagger\Bigr).
\end{align}
Thus
\begin{equation}
  L_{\gamma_\varphi}'
  \approx \sqrt{\gamma_\varphi/2}\Bigl(\sigma_z
         - 2\lambda(\sigma_+\tilde a + \sigma_-\tilde a^\dagger)\Bigr),
  \label{eq:SM3-LgammaPhi-prime}
\end{equation}
which generates dephasing-induced excitation exchange between atom and cavity.

\paragraph*{Approximation hierarchy.}
A fully rigorous treatment would keep all first-order corrections in
\eqref{eq:SM3-Lkappa-prime}, \eqref{eq:SM3-Lgamma1-prime},
\eqref{eq:SM3-LgammaPhi-prime} inside the dissipators
$\mathcal{D}[L_\kappa']$, $\mathcal{D}[L_{\gamma_1}']$,
$\mathcal{D}[L_{\gamma_\varphi}']$. In many dispersive experiments, however,
one operates in a parameter regime where
\begin{equation}
  \kappa\lambda^2 \ll \gamma_1,\qquad
  \gamma_\varphi\lambda^2 \ll \gamma_1,
  \label{eq:SM3-hierarchy}
\end{equation}
so that the cavity-induced corrections to the $\kappa$ and $\gamma_\varphi$
channels are negligible compared to the dominant atomic relaxation. In this
case one can safely neglect the $\lambda$ corrections in $L_\kappa'$ and
$L_{\gamma_\varphi}'$, while keeping the leading $O(\lambda^2)$ Purcell term
generated by $L_{\gamma_1}'$.

Concretely, inserting $L_{\gamma_1}'$ into $\mathcal{D}[L_{\gamma_1}']$ gives
\begin{equation}
  \mathcal{D}[L_{\gamma_1}']\rho'
  = \gamma_1\,\mathcal{D}[\sigma_-]\rho'
  - \gamma_1\lambda\,\mathcal{K}[\sigma_-,\sigma_z\tilde a]\,\rho'
  + \gamma_1\lambda^2\,\mathcal{D}[\sigma_z\tilde a]\,\rho',
  \label{eq:SM3-Dgamma1-expanded}
\end{equation}
where $\mathcal{K}$ denotes the interference superoperator. Under the usual
assumption that the interference term averages out in observables (e.g. on
timescales relevant for cavity dynamics), the dominant correction is the
second-order term, which yields an additional decay channel for the cavity via
atomic relaxation: the Purcell effect.

Within these approximations, the transformed master equation reads
\begin{equation}
\begin{split}
  \dot\rho'
  &\approx -i[H',\rho']
           + \kappa\,\mathcal{D}[\tilde a]\rho'
           + \gamma_1\,\mathcal{D}[\sigma_-]\rho' \\
  &\quad\;\; + \gamma_1\lambda^2\,\mathcal{D}[\sigma_z\tilde a]\rho'
           + \gamma_\varphi\,\mathcal{D}[\sigma_z]\rho',
\end{split}
  \label{eq:SM3-master-transformed}
\end{equation}
with $H'$ as in \eqref{eq:SM3-Hprime-dispersive}.

\subsection{Reduced cavity master equation and effective parameters}
\label{subsec:SM3-cavity-reduction}

We now trace out the atomic degree of freedom under the assumption that the
atom relaxes rapidly to a diagonal state with inversion
\begin{equation}
  \langle\sigma_z\rangle = p_e - p_g,
  \qquad p_e + p_g = 1,
\end{equation}
on the timescale of interest for the cavity dynamics. This is justified when
$\gamma_1$ is large compared to the effective cavity decay rate.

The coherent part of \eqref{eq:SM3-master-transformed} gives
\begin{align}
  \mathrm{Tr}_\text{atom}\bigl(-i[H',\rho']\bigr)
  &= -i\bigl[(\omega_c + \chi(B)\langle\sigma_z\rangle)
              \tilde a^\dagger\tilde a,\rho_\text{cav}\bigr],
\end{align}
where $\rho_\text{cav} := \mathrm{Tr}_\text{atom}\rho'$ is the reduced cavity
state. Thus the cavity resonance frequency is shifted to
\begin{equation}
  \omega_c'(B) := \omega_c + \chi(B)\,\langle\sigma_z\rangle.
  \label{eq:SM3-omega-c-prime}
\end{equation}

The cavity loss channel $\kappa\,\mathcal{D}[\tilde a]\rho'$ reduces to the
same form on the cavity. The Purcell term $\gamma_1\lambda^2\,\mathcal{D}[\sigma_z\tilde a]\rho'$
contributes an additional cavity dissipator $\propto\mathcal{D}[\tilde a]$
after tracing out the atom and inserting the inversion. Collecting these
contributions, one arrives at the reduced cavity master equation
\begin{equation}
\begin{aligned}
  \dot\rho_\text{cav}
  &= -i\bigl[\omega_c'(B)\,\tilde a^\dagger\tilde a,\rho_\text{cav}\bigr] \\
  &\quad + \kappa_\text{eff}(B)\,\mathcal{D}[\tilde a]\rho_\text{cav},
\end{aligned}
  \label{eq:SM3-master-cavity}
\end{equation}
with the field-dependent effective decay rate
\begin{equation}
  \kappa_\text{eff}(B) = \kappa + \gamma_1\left(\frac{g}{\Delta(B)}\right)^2.
  \label{eq:SM3-kappa-eff}
\end{equation}
The first term is the bare cavity loss, while the second is the Purcell
contribution arising from atomic relaxation in the dispersive regime.

Equations \eqref{eq:SM3-omega-c-prime} and \eqref{eq:SM3-kappa-eff} provide
the $B$-dependence of the GKSL data $H(B)$ and $F_k(B)$ required to compute
the QFIM blocks along the trajectory, in the notation of Appendix~\ref{app:formalism}.

\subsection{Cavity qubit dynamics and Bloch-vector representation}
\label{subsec:SM3-Bloch}

We restrict the cavity to the $\{\ket{0},\ket{1}\}$ subspace and prepare the
initial pure state
\begin{equation}
  \ket{\psi_0} = \cos\theta\,\ket{0} + \sin\theta\,\ket{1},
  \label{eq:SM3-psi0-cav}
\end{equation}
with $\theta\in[0,\pi/2]$. After tracing out the atom, the initial cavity
density matrix is
\begin{equation}
  \rho_\text{cav}(0)
  = \begin{pmatrix}
      \cos^2\theta & \cos\theta\sin\theta \\
      \cos\theta\sin\theta & \sin^2\theta
    \end{pmatrix},
  \label{eq:SM3-rho0-cav}
\end{equation}
in the ordered basis $\{\ket{0},\ket{1}\}$.

The master equation \eqref{eq:SM3-master-cavity} is that of a qubit subject to
a phase rotation at frequency $\omega_c'(B)$ and amplitude damping at rate
$\kappa_\text{eff}(B)$. Its exact solution is well-known and reads
\begin{equation}
\rho_\text{cav}(t;B)=
\begin{pmatrix}
  1 - \sin^2\theta\, e^{-\kappa_\text{eff} t} &
  \cos\theta\,\sin\theta\, e^{i\omega_c' t}\, e^{-\kappa_\text{eff} t/2} \\
  \cos\theta\,\sin\theta\, e^{-i\omega_c' t}\, e^{-\kappa_\text{eff} t/2} &
  \sin^2\theta\, e^{-\kappa_\text{eff} t}
\end{pmatrix},
\label{eq:SM3-rho-cav-t}
\end{equation}
where, for brevity, we write $\omega_c'=\omega_c'(B)$ and
$\kappa_\text{eff}=\kappa_\text{eff}(B)$.

Any qubit state can be written as
\begin{equation}
  \rho_\text{cav}(t;B)
  = \frac12\Bigl[\mathbb{I} + \vec S(t;B)\cdot\vec\sigma\Bigr],
  \label{eq:SM3-Bloch-decomp}
\end{equation}
with Pauli vector $\vec\sigma=(\sigma_x,\sigma_y,\sigma_z)$ and Bloch vector
$\vec S(t;B)=(S_x,S_y,S_z)$. From \eqref{eq:SM3-rho-cav-t} we obtain
\begin{equation}
\begin{aligned}
  S_x(t;B) &= 2\,\mathrm{Re}\,\rho_{01}(t;B)
           = \sin 2\theta\,\cos(\omega_c' t)\,e^{-\kappa_\text{eff} t/2},\\
  S_y(t;B) &= -2\,\mathrm{Im}\,\rho_{01}(t;B)
           = -\sin 2\theta\,\sin(\omega_c' t)\,e^{-\kappa_\text{eff} t/2},\\
  S_z(t;B) &= \rho_{11}(t;B) - \rho_{00}(t;B)
           = 2\sin^2\theta\,e^{-\kappa_\text{eff} t} - 1.
\end{aligned}
  \label{eq:SM3-S-components}
\end{equation}
The time derivatives are
\begin{equation}
\begin{aligned}
  \dot S_x(t;B)
  &= -\sin 2\theta\,e^{-\kappa_\text{eff} t/2}
     \Bigl[\omega_c'\sin(\omega_c' t)
           + \tfrac{\kappa_\text{eff}}{2}\cos(\omega_c' t)\Bigr],\\[1mm]
  \dot S_y(t;B)
  &= -\sin 2\theta\,e^{-\kappa_\text{eff} t/2}
     \Bigl[\omega_c'\cos(\omega_c' t)
           - \tfrac{\kappa_\text{eff}}{2}\sin(\omega_c' t)\Bigr],\\[1mm]
  \dot S_z(t;B)
  &= -2\kappa_\text{eff}\sin^2\theta\,e^{-\kappa_\text{eff} t}.
\end{aligned}
  \label{eq:SM3-Sdot-components}
\end{equation}
We will also need the squared norm and dot product
\begin{equation}
  |\dot{\vec S}(t;B)|^2
  = \dot S_x^2 + \dot S_y^2 + \dot S_z^2,
  \qquad
  \vec S\cdot\dot{\vec S} = S_x\dot S_x + S_y\dot S_y + S_z\dot S_z,
  \label{eq:SM3-Sdot-normdot}
\end{equation}
and
\begin{equation}
  |\vec S(t;B)|^2
  = S_x^2 + S_y^2 + S_z^2.
  \label{eq:SM3-S-norm}
\end{equation}

\subsection{Fidelity between initial and time-evolved cavity states}
\label{subsec:SM3-fidelity}

The initial cavity state is pure, $\rho_0 = \ket{\psi_0}\bra{\psi_0}$ with
$\ket{\psi_0}$ given by \eqref{eq:SM3-psi0-cav}. The fidelity between $\rho_0$
and $\rho_\text{cav}(t;B)$ is therefore
\begin{equation}
  F(\rho_0,\rho_\text{cav}(t;B))
  = \bra{\psi_0}\rho_\text{cav}(t;B)\ket{\psi_0}.
  \label{eq:SM3-fidelity-def}
\end{equation}
Writing $\ket{\psi_0} = (\cos\theta,\sin\theta)^\top$ in the
$\{\ket{0},\ket{1}\}$ basis and inserting \eqref{eq:SM3-rho-cav-t}, one finds
\begin{align}
  F(\rho_0,\rho_\text{cav}(t;B))
  &= \cos^2\theta\,\rho_{00}(t;B)
     + \sin^2\theta\,\rho_{11}(t;B)
     + 2\cos\theta\sin\theta\,\mathrm{Re}\,\rho_{01}(t;B) \nonumber\\
  &= \cos^2\theta\Bigl[1 - \sin^2\theta\,e^{-\kappa_\text{eff}t}\Bigr]
     + \sin^4\theta\,e^{-\kappa_\text{eff}t} \nonumber\\
  &\quad + 2\cos^2\theta\sin^2\theta\,e^{-\kappa_\text{eff}t/2}
           \cos(\omega_c' t).
\end{align}
Collecting terms,
\begin{equation}
\begin{aligned}
  F(\rho_0,\rho_\text{cav}(t;B))
  &= \cos^2\theta
     + \sin^4\theta\,e^{-\kappa_\text{eff}t}
     - \cos^2\theta\sin^2\theta\,e^{-\kappa_\text{eff}t} \\
  &\quad
     + 2\cos^2\theta\sin^2\theta\,e^{-\kappa_\text{eff}t/2}
       \cos(\omega_c' t).
\end{aligned}
  \label{eq:SM3-fidelity-explicit}
\end{equation}
This expression will enter the numerator of the projected QSL.

\subsection{QFI for time and field; mixed element}
\label{subsec:SM3-QFIM}

For a general mixed qubit state with Bloch vector $\vec S(x)$ depending on a
real parameter $x$, the SLD QFI is given by the standard formula
\begin{equation}
  F_Q(x)
  = \bigl|\partial_x\vec S\bigr|^2
    + \frac{\bigl(\vec S\cdot\partial_x\vec S\bigr)^2}
           {1-|\vec S|^2},
  \qquad |\vec S|<1,
  \label{eq:SM3-QFI-Bloch}
\end{equation}
with the pure-state limit obtained by continuity as $|\vec S|\to 1$.

\paragraph*{Time QFI.}
Taking $x=t$ with $B$ fixed, we write
\begin{equation}
  F_{tt}(t;B)
  := F_Q(t)
  = |\dot{\vec S}(t;B)|^2
    + \frac{\bigl[\vec S(t;B)\cdot\dot{\vec S}(t;B)\bigr]^2}
           {1-|\vec S(t;B)|^2},
  \label{eq:SM3-Ftt-open}
\end{equation}
with $\vec S$ and $\dot{\vec S}$ given by
\eqref{eq:SM3-S-components}--\eqref{eq:SM3-Sdot-components}. This is the
physical Bures speed squared (up to the factor $1/4$) along the open-system
trajectory for fixed field.

\paragraph*{Field QFI.}
The field enters through the effective frequency $\omega_c'(B)$ and decay
rate $\kappa_\text{eff}(B)$. It is convenient to denote
\begin{equation}
  \Omega(B) := \omega_c'(B),\qquad
  \Gamma(B) := \kappa_\text{eff}(B),
\end{equation}
and their derivatives
\begin{equation}
  \Omega'(B) := \partial_B\Omega(B),\qquad
  \Gamma'(B) := \partial_B\Gamma(B).
\end{equation}
From \eqref{eq:SM3-omega-c-prime} and \eqref{eq:SM3-kappa-eff},
\begin{align}
  \Omega'(B)
  &= \chi'(B)\,\langle\sigma_z\rangle
   = \langle\sigma_z\rangle\,\partial_B\Bigl(\frac{g^2}{\Delta(B)}\Bigr)
   = -\,\langle\sigma_z\rangle\,\frac{g^2\Delta'(B)}{\Delta(B)^2},\\
  \Gamma'(B)
  &= \partial_B\Bigl[\kappa + \gamma_1\Bigl(\frac{g}{\Delta(B)}\Bigr)^2\Bigr]
   = -\,2\gamma_1\,\frac{g^2\Delta'(B)}{\Delta(B)^3},
\end{align}
with $\Delta'(B) = -\omega'_a{}'(B)$ from \eqref{eq:SM3-omega-a-prime}.

Differentiating \eqref{eq:SM3-S-components} with respect to $B$ yields
\begin{equation}
\begin{aligned}
  \partial_B S_x
  &= -\sin 2\theta\,e^{-\Gamma t/2}
     \Bigl[t\,\Omega'\sin(\Omega t)
           + \tfrac{t}{2}\Gamma'\cos(\Omega t)\Bigr],\\[1mm]
  \partial_B S_y
  &= -\sin 2\theta\,e^{-\Gamma t/2}
     \Bigl[t\,\Omega'\cos(\Omega t)
           - \tfrac{t}{2}\Gamma'\sin(\Omega t)\Bigr],\\[1mm]
  \partial_B S_z
  &= -2\sin^2\theta\,t\,\Gamma'\,e^{-\Gamma t},
\end{aligned}
  \label{eq:SM3-SB-components}
\end{equation}
where again $\Omega=\Omega(B)$ and $\Gamma=\Gamma(B)$.

The field QFI is then
\begin{equation}
  F_{BB}(t;B)
  := F_Q(B;t)
  = |\partial_B\vec S(t;B)|^2
    + \frac{\bigl[\vec S(t;B)\cdot\partial_B\vec S(t;B)\bigr]^2}
           {1-|\vec S(t;B)|^2},
  \label{eq:SM3-FBB-open}
\end{equation}
with $|\partial_B\vec S|^2$ and $\vec S\cdot\partial_B\vec S$ evaluated using
\eqref{eq:SM3-S-components} and \eqref{eq:SM3-SB-components}. This is the
nuisance information needed in the Schur complement.

\paragraph*{Mixed element.}
The mixed QFI element $F_{tB}$ can be written in Bloch form as
\begin{equation}
  F_{tB}(t;B)
  = \dot{\vec S}(t;B)\cdot\partial_B\vec S(t;B)
    + \frac{\bigl[\vec S(t;B)\cdot\dot{\vec S}(t;B)\bigr]
           \bigl[\vec S(t;B)\cdot\partial_B\vec S(t;B)\bigr]}
           {1-|\vec S(t;B)|^2}.
  \label{eq:SM3-FtB-open}
\end{equation}
Equations \eqref{eq:SM3-Ftt-open}, \eqref{eq:SM3-FBB-open}, and
\eqref{eq:SM3-FtB-open} provide the full $2\times2$ QFIM in the parameter pair
$(t,B)$ for the open, dispersive JC sensor.

\subsection{Schur-complement information and projected QSL}
\label{subsec:SM3-projected}

Treating $t$ as the parameter of interest and $B$ as a nuisance, the effective
information for time is given by the one-dimensional Schur complement
\begin{equation}
  F_{\mathrm{eff}}(t;B)
  = F_{tt}(t;B) - \frac{F_{tB}(t;B)^2}{F_{BB}(t;B)},
  \label{eq:SM3-Feff-open}
\end{equation}
with the understanding that $F_{BB}^{-1}$ is interpreted as the Moore--Penrose
pseudoinverse (and $F_{\mathrm{eff}}=F_{tt}$) whenever $F_{BB}=0$.
By positive semidefiniteness of the QFIM, $F_{\mathrm{eff}}\ge 0$, and
$F_{\mathrm{eff}}\le F_{tt}$ with equality if and only if $F_{tB}=0$.

The instantaneous projected (quotient) speed is
\begin{equation}
  v_{\mathrm{quo}}(t;B) = \frac12\sqrt{F_{\mathrm{eff}}(t;B)},
  \label{eq:SM3-vquo-open}
\end{equation}
and its time average over the interrogation interval $[0,\tau]$ is

\begin{equation}
  \bar v_{\mathrm{quo}}(B)
  = \frac{1}{\tau}\int_0^\tau \frac12\sqrt{F_{\mathrm{eff}}(t;B)}\,dt.
  \label{eq:SM3-vquo-open-avg}
\end{equation}
For a fixed field $B$, this reduces to the standard QSL denominator, but with
the profiled speed rather than the physical one.

Because $B$ is a nuisance parameter that may be re-tuned between experimental
runs within an admissible set $\mathcal{B}$, the numerator must live on the
quotient manifold in which states that differ only by a change in $B$ are
identified. The relevant Bures angle is therefore
\begin{equation}
  \Theta_{\mathrm{quo}}(\tau)
  := \arccos\!\Bigl(
      \sup_{B\in\mathcal B}
      \sqrt{F(\rho_0,\rho_\text{cav}(\tau;B))}\Bigr),
  \label{eq:SM3-theta-quo-open-def}
\end{equation}
where $F(\rho_0,\rho_\text{cav}(\tau;B))$ is given explicitly in
\eqref{eq:SM3-fidelity-explicit}.

The quotient angle is obtained by optimizing the explicit fidelity
\eqref{eq:SM3-fidelity-explicit} over the admissible calibration window
$\mathcal B$. Define
\begin{equation}
\begin{aligned}
  F(B;\tau)
  :=\;& F\!\bigl(\rho_0,\rho_{\mathrm{cav}}(\tau;B)\bigr) \\
  =\;& \cos^2\theta
      + \sin^4\theta\,e^{-\kappa_\text{eff}(B)\tau}
      - \cos^2\theta\sin^2\theta\,e^{-\kappa_\text{eff}(B)\tau} \\
   &\quad
      + 2\cos^2\theta\sin^2\theta\,
        e^{-\kappa_\text{eff}(B)\tau/2}
        \cos\!\bigl(\omega_c'(B)\tau\bigr).
\end{aligned}
  \label{eq:SM3-f-open-aux}
\end{equation}
Then Eq.~\eqref{eq:SM3-theta-quo-open-def} becomes
\begin{equation}
  \Theta_{\mathrm{quo}}(\tau)
  =
  \arccos\!\left(
    \sup_{B\in\mathcal B}\sqrt{F(B;\tau)}
  \right).
  \label{eq:SM3-theta-quo-open}
\end{equation}

Combining the quotient distance and the projected speed, the nuisance-profiled
QSL for a fixed realized field value $B$ in the open JC sensor reads
\begin{equation}
  \tau \;\ge\;
  \frac{\Theta_{\mathrm{quo}}(\tau)}
       {\displaystyle
        \frac{1}{\tau}\int_0^\tau
          \frac12\sqrt{F_{tt}(t;B) - \frac{F_{tB}(t;B)^2}{F_{BB}(t;B)}}\,dt},
  \label{eq:SM3-QSL-open}
\end{equation}

with $F_{tt}$, $F_{BB}$, and $F_{tB}$ given by
\eqref{eq:SM3-Ftt-open}, \eqref{eq:SM3-FBB-open}, and
\eqref{eq:SM3-FtB-open}, and $\Theta_{\mathrm{quo}}(\tau)$ given by
\eqref{eq:SM3-theta-quo-open}.



   If the dynamics is insensitive to $B$ over
  the admissible range (e.g.\ $\Omega'(B)\equiv 0$ and $\Gamma'(B)\equiv 0$ so
  that $\partial_B\vec S\equiv 0$), then $F_{BB}=F_{tB}=0$ and
  $F_{\mathrm{eff}}=F_{tt}$. The quotient angle reduces to the physical Bures
  angle $L(\rho_0,\rho_\text{cav}(\tau;B))$ for any $B$, and
  \eqref{eq:SM3-QSL-open} collapses to the standard open-system QSL. {If $F_{tB}\equiv 0$ while $F_{BB}>0$, then $F_{\mathrm{eff}}=F_{tt}$, so the
projected and physical QSL denominators coincide for the same realized field
value $B_0$. The numerator remains profiled via
$\Theta_{\mathrm{quo}}(\tau)\le
\arccos\sqrt{F(\rho_0,\rho_\text{cav}(\tau;B_0))}$, because the quotient angle
is obtained by maximizing the fidelity over the admissible calibration window
before applying the decreasing function $\arccos$. The projected QSL lower bound
is therefore weaker than or equal to the corresponding physical QSL lower bound.}  In the limit $\kappa_\text{eff}\tau\ll 1$, $\Gamma'(B)\tau\ll 1$, and with detuning tolerance chosen such that the coherent dynamics dominates over decay, the  open-system QFIM and projected QSL smoothly approach their closed-system counterparts of Appendix~\ref{app:JC-unitary})

\end{widetext}

\end{document}